\documentclass[prb]{revtex4-1} 

\usepackage{amsmath} 
\usepackage{amsfonts}
\usepackage{graphicx}
\usepackage{color}
\usepackage{soul}

\graphicspath{{Figures/}}

\definecolor{dgreen}{rgb}{0,0.5,0}
\definecolor{palered}{rgb}{1,0.5,0.5}
\definecolor{purple}{rgb}{1,0,1}

\newcommand{\showfigures}[1]{#1}

\setstcolor{red}

\newcommand{\modifxE}[1]{\textcolor{black}{#1}} 

\newcommand{\beginsupplement}{
      \setcounter{section}{0}
        \renewcommand{\thesection}{S\arabic{section}}%
        \setcounter{table}{0}
        \renewcommand{\thetable}{S\arabic{table}}%
        \setcounter{figure}{0}
        \renewcommand{\thefigure}{S\arabic{figure}}%
        \setcounter{page}{1}
     
      \clearpage
     }

\begin{document}


\title{Inference uncertainty about an aircraft crash}

\author{Fran\c{c}ois Graner}
\email{francois.graner@u-paris.fr}
\affiliation{Universit\'e Paris Cit\'e, CNRS, Mati\`ere et Syst\`emes Complexes, Paris, France.}

\author{Stefano Matthias Panebianco}
\email{stefano.panebianco@cea.fr}
\affiliation{Universit\'e Paris-Saclay, Centre d'Etudes de Saclay (CEA), IRFU, D\'epartement de Physique Nucl\'eaire (DPhN), Saclay, France}

\date{\today}

\begin{abstract}
Problem-based learning benefits from situations taken from real life, which usually stimulate student interest. 
In this paper, we examine the shooting down of the Rwandan president's aircraft
on April 6th, 1994.
We discuss methods to infer information about
 the location from which the missile was launched,  its trajectory and type,  where the aircraft was struck and its trajectory during the fall.
To this end, we developed a physics-based analysis based on expert reports, witness statements, and other publicly available information, as interpreted by our calculations. 
The analysis is designed to be understandable using undergraduate-level physics.
The uncertainty of each result is discussed and propagated to ensure a proper assessment of the hypotheses and a traceability of their consequences. 
Such approach  encourages the students to exercise their critical mind and teaches inference methods that are routinely used in physics research. 
In addition, it illustrates the importance and limits of scientific expertise during a judiciary process.
\end{abstract}

\maketitle

\section{Introduction}
\label{sec:introduction}

\subsection{Motivation}
 \label{sec:motivation}
  
Daily-life phenomena constitute a rich source of problem-based inquiry in physics, as they require students to construct models, formulate and test hypotheses, and critically assess the reliability and uncertainty of their results. Working on such open-ended problems requires students to translate qualitative observations into quantitative equations, reconcile incomplete or inconsistent data, and evaluate the degree to which available information constrains their conclusions. This approach is particularly evident in forensic physics, a domain often introduced at the undergraduate level, where real-world case studies significantly enhance student engagement and motivation. 

In this work,
we examine a historically significant
case, the 1994 shooting of the Rwandan presidential aircraft.
 \label{sec:event}
The Falcon 50 aircraft registered as 9XR-NN 
was shot down in Rwanda on April 6th, 1994.
The aircraft's owner, Juv\'enal Habyarimana, President of Rwanda, and his guest Cyprien Ntaryamira, President of Burundi, had spent the day in Dar es Salam, Tanzania, for a regional summit. 
In the evening, with 7 other passengers and 3 crew members, their aircraft was
nearing  Rwanda capital's Kigali  when it was shot by a missile.
All 12 people onboard died in the crash. 
Within a day, Rwandan army units and extremist Hutu militias began to perpetrate the Tutsi genocide on a large scale.
Using undergraduate-level physics and the publicly available information, we examine  the origin of the launch  and the location of the missile strike.

\subsection{Sources and data}
 \label{sec:facts}

\subsubsection{Sources and data treatment}
 \label{sec:sources}
 
French soldiers investigated the aircraft during the night of the crash, and the French judiciary conducted witness hearings~\cite{bruguiere}.
In spring and summer 1994, Belgian military investigators questioned witnesses, and on August 1st  two of them examined the aircraft fragments remaining at the site~\cite{belgian_investigators}.
On April 10th, 2002 a French aviation accident expert issued an analysis of some of the air traffic control recordings~\cite{bande}.
A seven-member Rwandan commission  heard 577 witnesses from February 2008 to February 2009, 
commissionned two UK military experts (ammunitions and forensic analysis) 
who  issued their report~\cite{cranfield} on February 27, 2009,
and then  published its own final report on April 20, 2009~\cite{mutsinzi}
along with a brief video synopsis~\cite{mutsinzi_video}.
The report of a group of six French experts in  aviation, ballistics, weapons, explosives, land survey~\cite{oosterlinck} and acoustics~\cite{acoustic}
 was provided on January 5, 2012.

 \label{sec:approach}

We  consulted several aeronautic or military websites as of  early 2024 (except when noted otherwise); and also expert reports available  online regarding aeronautics, weapons, ballistics or radio communications, 
as well as on-site measurements and tens of witness accounts. 
Exploiting the sources involves a few technical details (Section~\ref{sec:supp_aero_units}).
We converted angles to degrees, referred them to the geographic North. We defined  counterclockwise angles as positive, following the trigonometric convention  
(as opposed to the compass convention, in which clockwise angles are counted as positive).  
 \label{sec:units}
We converted other quantitative data to SI units.
We tried to accurately determine, combine and trace the uncertainty sources.
\label{sec:rounding}
In order to avoid propagation of rounding errors,
we kept digits in excess throughout all calculations.
For legibility and consistency we should have rounded the values and uncertainties, 
but then calculations made with these rounded values might have significantly differed from the calculations we have actually made (see Section~\ref{sec:sensitivity} and Table~\ref{table:results}).  

\subsubsection{Airport data}
 \label{sec:airport}
 
Kigali  airport
is located in Kanombe, in the eastern suburbs of  Kigali.
We use a Cartesian coordinate system with the $z$-axis vertical. We define $z = 0$ at sea level, so the point labeled $O$ in Fig.~\ref{fig:interception_position}, called ``threshold", at the east end of runway 28, is at $z_O = 1486$~m. 
The runway is on the $x$ axis, with $x$ pointing West to East. The $y$ axis is the horizontal axis perpendicular to the runway, oriented from South to North.
According to  the standard instrument approach procedure, an aircraft
is expected to follow a well-defined trajectory (Fig.~\ref{fig:schema_trajectories})
 in the vertical $Oxz$ plane,
at an
angle $\gamma = 3.00^\circ$  from the horizontal,
called the ``glide". The equation for the trajectory then is:
\begin{eqnarray}
y &=& 0 
\label{eq:glide_slope_a} \\
z &=& 
z_O+  x \; \tan \gamma
=1486 + 0.052 \; x.
\label{eq:glide_slope_b}
\end{eqnarray} 
This prescribed trajectory starts
at $x_G=14446$~m  (Ref.~\cite{cranfield} p.~106),
  ends  at the runway threshold $O$, and its angular uncertainty (seen from $O$) both in  $y$ and in $z-z_O$ is 
 $\pm 0.5^\circ$~\cite{FSF,CANSO}.

Hence we retain an uncertainty $\pm 8.7\; 10^{-3}\; x$ on the value of both $y$ and $z-z_O$.
We refer the reader to Table~\ref{table:results}  which is a guide through the physical parameters.
\subsubsection{Crash data}
 \label{sec:interception_description}

According to the  available recorded conversations between the pilots and the air traffic controllers~\cite{bande},
 the crew announced at $t_A=20$:21:27 that the aircraft was at point $A$ with $x_A = 37040$~m and $z_A=3657$~m.

At 20:21:42, it requested a direct approach to  runway 28.  
The controller informed the aircraft (Ref.~\cite{bande} p. 11) that the wind speed was 2.06 m/s, the temperature was 19$^\circ$C, and that ceiling and visibility were satisfactory.
\label{sec:CAVOK}
More than two hours after sunset, it was  fully dark. 

Around 20:26, the aircraft had already begun its descent toward the airport.
Most witnesses saw  two successive luminous lines coming from the southern side of the aircraft: the first one passed close to the aircraft and missed it; the second one intersected its trajectory. It what follows (unless explicitly stated) we have assumed that it is correct.  The witness accounts do not allow one to determine what actually happened, in particular whether the missile hit the aircraft or exploded nearby.
Section~\ref{sec:missile_number} discusses whether the luminous streaks indeed corresponded to missile trajectories, and whether there were two of them.

The location $E$ and time $t_E$ of the aircraft-missile encounter (aircraft interception  by the missile)  are not exactly known.
A signal was detected on the distress frequency during 22 seconds, between  20:25:50 and  20:26:12  (Ref.~\cite{bande} p.~10-13).
The French aviation accident expert assumed this whole time interval corresponded to the time of the crash (Ref.~\cite{bande} p.~12), meaning he could not distinguish between the time of the interception of the aircraft by the missile,  $t_E$, and that of the aircraft's impact on the ground, $t_I$, due to the unusual nature of the distress signal.
Given the lack of information, we assumed that  $t_E$  and  $t_I$ were  close to each other,  both being around 20:26:01~$\pm 11$~s. We also assumed that the order of magnitude of their difference $D=t_I-t_E$ was  ten seconds (see Sections~\ref{sec:free_fall}, \ref{sec:longitudinal_fall}). 

\begin{figure}[t]
(a) \hfill ~ \\
\hfill ~ \\
\showfigures{\includegraphics[width=0.99\textwidth]{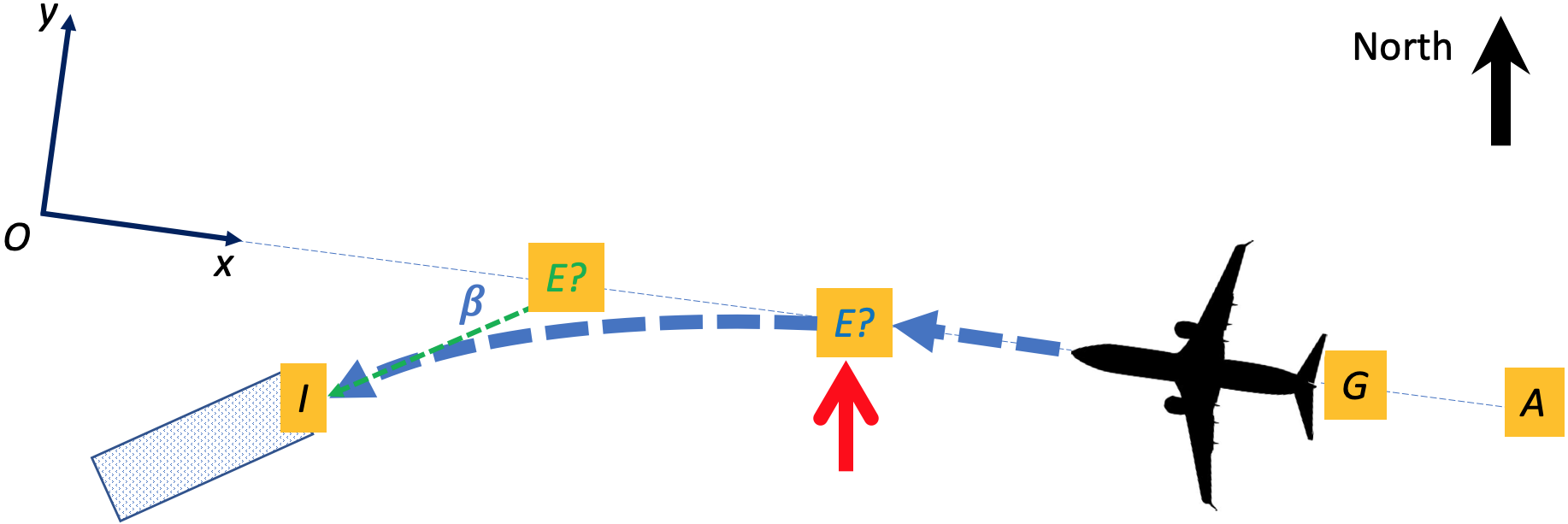}}\\
\hfill ~ \\
(b)  \hfill ~ \\
\hfill ~ \\
\showfigures{\includegraphics[width=0.99\textwidth]{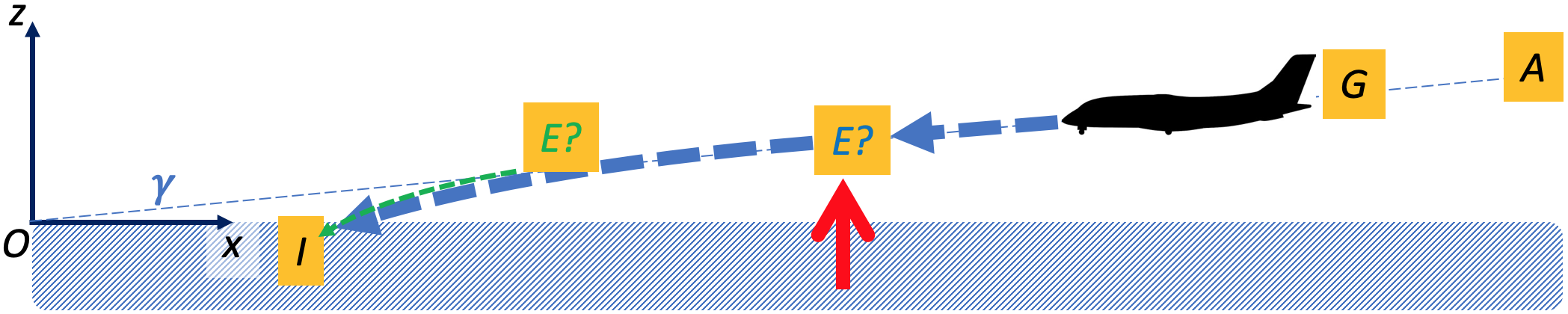}}\\
\caption{{\bf Trajectory of the aircraft before and after the missile-aircraft interception at point $E$:}
(a) Top view. (b) Side view in the $Oxz$ plane. Not drawn to scale.
 We know aircraft's positions at point $A$ when the crew communicated with the air traffic controllers,  at point $G$ when it reached the glide, and at point $I$ when it impacted the ground. We also know the position of the debris (located in the dotted rectangle around $I$), the lateral deviation $\beta$ from the glide at impact, and the vertical descent angle $\gamma$. We do not know the position of point $E$.
Trajectory from $E$ to $I$ is drawn according either to the free fall hypothesis discussed in Section~\ref{sec:free_fall} (thin dashed green arrows), or to the  progressive deviation hypothesis discussed in Section~\ref{sec:longitudinal_fall}
(thick blue dashed arrows).
The momentum of a missile shot from the ground and directed from South to North towards the aircraft is represented by thick red solid arrows. 
}
\label{fig:schema_trajectories}
\end{figure}

The aircraft impact position $I$ on the ground  is known. In 1994, the impact trace was visible as a 8 m diameter circle. Its position has been determined with a few meters uncertainty by Belgian investigators, along with that of several aircraft fragments dispersed on the ground over a rectangle {(Fig.~\ref{fig:schema_trajectories}a)} of 145~m $\times$ 20~m (Ref.~\cite{belgian_investigators} p.~5). 
In 2012, French investigators  combined their own geometrical measurements with the 1994 Belgian report to determine the coordinates of  the impact position $I$ within a few meters:  $x_I = 2160$~m, $y_I = -100$~m and  $z_I=1410$~m  (Ref.~\cite{oosterlinck} p.~188, 193). 

\section{Interception position}
 \label{sec:fall}

The exact position $E$ at which the missile intercepted the aircraft is not known.
To determine it, we first need to calculate the aircraft's longitudinal speed $\dot{x}$ (hereafter called ``speed") and its direction both at $t_E$ and $t_I$  (Sections~\ref{sec:speed_position},~\ref{sec:aircraft_mass_before},~\ref{sec:lateral_fall}). Then, we will reverse analyse  the aircraft trajectory and its velocity variation between $E$ and $I$ (Sections~\ref{sec:free_fall}, \ref{sec:longitudinal_fall}).  This will enable us to infer the location of $E$.  In all what follows, the dot indicates time derivative.

\subsection{Aircraft speed before the interception}
 \label{sec:speed_position}
 
The speed  between crew announcement $A$ and impact $I$ is not constant.
In addition, the deceleration between the interception and the impact has no reason to be equal to the prescribed acceleration.
However, since the trajectory between $E$ and $I$ is only a tiny portion of the trajectory between $A$ and $I$, 
and since the aircraft inertia implied that ${\dot{x}}$ only varied by a few percents between $E$ and $I$,
to fix the ideas  we can roughly extrapolate the prescribed acceleration after the interception until $I$, which probably introduces only a small error on  ${\dot{x}}_I$.
We can thus estimate the average speed  between crew announcement $A$ and impact $I$, and the position as a function of time, as follows.

For simplicity we temporarily assimilate $E$ and $I$ and assume $t_I= 20$:26:01~$\pm$~11~s.
In a time $t_I-t_A=274 \pm 11$~s, the aircraft covered 
a distance 
$34.9  \pm 1.8$~km. 
The average speed therefore is $127 \pm 8$~m/s,
assuming the uncertainties on distance and time are independent. This  value is consistent with the approach speed of a Falcon 50~\cite{eurocontrol}.
 We assume that the pilot was planning a landing in accordance with the Aircraft Performance Database~\cite{eurocontrol}, which calls for a roughly constant deceleration that reaches a position
 ${x=1100}$~m east of the landing strip with a speed of 67~m/s. 

 We know the impact position $x_I$ and time $t_I$.
The two unknowns we are looking for are ${\dot{x}}_A$ and the acceleration $a={\ddot{x}}$ along the $x$ direction (note that ${\dot{x}}<0$ hence $a>0$).
To estimate  $a$'s order of magnitude 
we express the speed and position as a function of time: 
\begin{eqnarray}
{\dot{x}}(t)&=&a(t-t_A) + {\dot{x}}_A, 
\label{eq:speed_position_vs_time_a} \\
x(t) &=& \frac{a}{2} (t-t_A)^2 + {\dot{x}}_A(t-t_A) + x_A.
\label{eq:speed_position_vs_time_b}
\end{eqnarray}
By eliminating $t-t_A=({\dot{x}}(t)-{\dot{x}}_A)/a$ 
 the velocity can be expressed as a function of distance: 
\begin{equation}
{\dot{x}}(x) = - \left[{\dot{x}}_A^2 - 2 a (x_A-x)\right]^{1/2}.
\label{eq:speed_profile}
\end{equation}

Eliminating $a$ from the equations representing the two constraints yields a quadratic equation on ${\dot{x}}_A$, with two  negative solutions ${\dot{x}}_A=-342.5$~m/s or ${\dot{x}}_A=-181.5$~m/s.
While the former value is irrealistic, and beyond the reach of a Falcon 50, the latter
 is consistent with aircraft data~\cite{eurocontrol}.
It yields $a=0.397$~m/s$^2$, which is positive as expected. 
By propagating the uncertainties on the quadratic equation, 
we calculate that uncertainties  are around
$\pm 10$~m/s for speed and 30\%
\label{sec:speed_at_impact}
To fix the ideas, if we extrapolate Eq.~(\ref{eq:speed_profile}) until the impact, this determine the speed at impact, ${\dot{x}}_I=
-72.7$~m/s.
We recall that the deceleration between the interception and the impact has no reason to be equal to the prescribed acceleration, but since Eq.~(\ref{eq:speed_profile})
corresponds to a much longer trajectory, and since the aircraft inertia implied that ${\dot{x}}$ only varied by a few percents between $E$ and $I$,
 a rough extrapolation of the prescribed speed  after the interception until $I$ probably introduces only a small error on  both  ${\dot{x}}_E$ and ${\dot{x}}_I$.
 
As a by-product of Eqs.~(\ref{eq:speed_position_vs_time_a}-\ref{eq:speed_profile}) we also obtain that the aircraft reached the glide ($x_G=144{46}$~m)
at time $t_G=t_A+ 149$~s, i.e.  20:23:56, which, as expected, is later than the  crew declaration (20:21:42)  of its intention to reach it  (Ref.~\cite{bande} p.~12). 
It then flew at ${\dot{x}}_G=-122.5\pm 10$~m/s,
a  value which is consistent with aircraft data\cite{eurocontrol} although rather high  (Ref.~\cite{cranfield} p.~106). 
This might mean that our constant deceleration hypothesis 
 should be slightly corrected, 
with more deceleration occurring before reaching the glide (from $A$ to $G$) and less deceleration later (from $G$ to $E$ or $I$). 
In the absence of more definitive information, we do not develop the model further and retain the simplest hypothesis.

\subsection{Continuity of aircraft speed at interception}
 \label{sec:aircraft_mass_before}

\label{sec:momentum_transfer}
To determine the velocity  vector after the interception   as a function of time or position, it is first useful to determine 
the extent to which the aircraft's velocity vector was affected by the interception by the missile at $t_E$.
This change may arise from a number of factors.
If the missile struck the aircraft, momentum transfer from the missile could have contributed to the change in the aircraft's velocity vector. 
This change would then have been oriented from south to north and from low to high altitude (see the thick red solid arrows in Fig.~\ref{fig:schema_trajectories}). 
We will discuss the effect of such collision on the angular velocity in Section~\ref{sec:lateral_fall}. 
However, an upper bound for the magnitude of the change in momentum can already be estimated.
The order of magnitude of the aircraft's mass $M$, including fuel and passengers, was greater than 10$^4$~kg
(when empty, its mass is  9163~kg~\cite{wp_falcon}). 
The missile considered here  has a mass $m$ on the order of at most 20~kg and a speed of at most 1000~m/s~\cite{oosterlinck}, whatever its type (discussed in Section~\ref{sec:missile_type}).
The mass ratio $m/M$ was therefore at most of 2~10$^{-3}$, so that the change in the aircraft's velocity vector
could be at most on the order of 2~m/s. 
As a consequence, the direct momentum transfer from the missile cannot have caused a significant discontinuity in the aircraft's velocity vector.

A similar calculation can estimate the momentum transfer due to the warhead explosion.
The different missile types (discussed in Section~\ref{sec:missile_type}) have a similar order of magnitude of released energy. 
To estimate it we consider for instance the Mistral's explosive charge, three~kg of hexogen-tolite, which projects around 1500 tungsten beads~\cite{mindef_mistral}. 
Hexogen-tolite yields 
a detonation velocity $v_d = 8750$~m/s
and releases
1.6 times more  energy per kg than  TNT  (4.184~$10^6$ J)~\cite{wp_RDX}. 
The energy released upon explosion was thus of order of $E_{\rm expl}\simeq 2 \; 10^7$ J. Part of it was converted into kinetic energy $E_c$ of the air (shock wave), 
fragmenting the missile and projecting tungsten beads (of total mass $m_t$). The momentum of the tungsten beads would then be, at most, of $m_tv_d=2E_{\rm expl}/v_d=4.6 \; 10^4$~kg~m/s.
Hence, if the momentum transfer had been perfect and  unidirectional, the aircraft's velocity could have changed of order of 0.5~m/s; in practice it was probably much less.
\label{sec:impulse}

An evasive maneuver could also have been possible~\cite{RapportEvitement}:
\label{evasive_left}
 If the pilot saw the first missile (which missed the aircraft), he could have tried to suddenly change its trajectory to avoid the second missile. Then, the change in the aircraft's velocity would have been delayed both by the pilot's reaction time (typically on the order of $10^{-1}$~s~\cite{Hick}) and by the aircraft's reaction time (at least of the same order of magnitude). The time interval between missile launches was between a fraction of second and many seconds, and the duration of the missile trajectories was of order of a few seconds (Section~\ref{sec:missile}).  If the pilot saw the launch of the first missile and had time to react before the interception by the second one, he could have modified significantly the aircraft direction. But we can speculate that if he had seen a missile coming from his left, he would have performed a maneuver toward his right, which is not compatible with the observed deviation toward his left~\cite{RapportEvitement} (Section~\ref{deviation_left}). If the pilot had only one second to react,  the velocity's modulus could then have varied by a few m/s (Figs. 31, 34 of Ref.~\cite{Akdag_master}), which is negligible.

\subsection{Lateral deviation after interception}
\label{sec:lateral_fall}

While Section~\ref{sec:momentum_transfer} was devoted to the discontinuous change in direction {\it at}  the interception by the missile, we now turn to the change in direction {\it after} the interception, during the fall.
Because of  the damages, the direction of the aircraft body symmetry axis and that of the aircraft center-of-mass velocity vector are not necessarily parallel. We should study them as independent unknown variables. The first known information is the locations of impact $I$, at 100~m South of its planned trajectory.
The second known information is the location of debris.
Drawing a line through the 1994 Belgian report on fragment positions (Ref.~\cite{belgian_investigators} p.~5) combined with their own
measurements, 
French investigators have  determined that the aircraft bounced on the ground at an angle of $13.7^\circ$ from its planned trajectory (Ref.~\cite{oosterlinck} p.~188, 189), i.e. at an angle of  $193.7^\circ$ from the $x$ axis.
Based on the 
size of the rectangle where the debris were found (length 145~m and half-width 10~m, Section~\ref{sec:interception_description}) 
we estimate the uncertainty on this angle  to be $\pm \arctan(10/145) = \pm 4^\circ$.
\label{sec:orientation_fragment}
In addition, we have not detected any asymmetry of the fragment dispersion, 
with respect to the aircraft bounce direction, hence the fragment locations probably reflect the direction of  the aircraft sliding. 
This suggests that, at the time of impact, the aircraft body and velocity vector were parallel.
In turn, this suggests that the aircraft body and velocity vector turned left at the same speed during the fall, until both had   deviated laterally by $\beta =  13.7\pm 4^\circ$ and then kept their direction during the sliding.
In the absence of more definitive information on how the impact could have affected the aircraft body direction, we do not develop the model further.

The deviation towards the left could in principle be explained by an initial discontinuity at the time of interception.
\label{deviation_left}
Other explanations which could have caused the observed deviation towards the left include: larger damages by the missile on the left engine than to the right one, and/or a progressive loss of lift on the left wing due to  damages.  
The latter explanation would suggest that the  wing plane became tilted to the left, with respect to horizontal (``roll"). This is compatible with the observations by  Belgian investigators that only the left side of the plane had directly hit the ground, and that the left wing was much damaged while the right one was intact (Ref.~\cite{belgian_investigators} p.~3).

Quantitatively,   the aircraft size is 18.5~m in both length and width~\cite{wp_falcon}, and its mass is known. We can easily estimate the inertia matrix as that of a cross: it is independent on the material composition or thickness, and largely independent of the exact geometry and spatial density heterogeneities.
We obtain that the main inertia moment is of order of 2.85~10$^5$~kg~m$^2$. The instantaneous moment transfer (or ``impulse") 
of about 4.6~10$^4$~kg~m/s directly from the missile or from its explosion (see Section~\ref{sec:impulse}) off-centered by 9~m could create an angular velocity after interception with an order of magnitude of up to 1.5~$10^{-1}$~s$^{-1}$. This would be enough to make the aircraft turn progressively by 13.7$^\circ$ during its fall.
\subsection{Estimates of interception position}
\label{sec:longitudinal_fall}

To find the position $x_E$ at which the missile intercepted the aircraft, we 
have to connect the aircraft's trajectory after the interception with the planned trajectory
(Fig.~\ref{fig:schema_trajectories}a).
To go beyond the  free fall model (Section~\ref{sec:free_fall}), we  suggest three alternate determinations for $x_E$, one based on geometry, one based on witness accounts, and one mixing both (Fig.~\ref{fig:interception_position}).

(i) The first determination is based on  the horizontal trajectory, $y(x)$. 
It starts from $E$, on the $x-$axis ($y_E=0$), with a tangent parallel to the $x-$axis (Fig.~\ref{fig:schema_trajectories}a) if we neglect the discontinuity in velocity direction due to the interception by the missile (Section~\ref{sec:momentum_transfer}). It ends on the ground impact  $(x_I,y_I) = (2160, -100)$ with a tangent angle   $(dy/dx)_I=\tan {\beta}$. It is thus curved. For simplicity we assume that its curvature is constant or, almost equivalently, that it is a  portion of parabola, which makes the calculation easier.
One finds that 
$x_E - x_I = 2 y_I / \tan {\beta}   = 820$~m. 
This is twice the value it would have in a free fall and
yields $x_E = 2160 + 820 = 2980$~m. The uncertainty on $\beta$ yields an asymmetric   confidence interval on $\tan {\beta}$ and we find 
$x_E$
between  
2666 and 3318~m.

(ii) The second determination is based on  witness accounts. 
Witnesses  at the airport had a good visibility of the aircraft during its descent, but their line of sight was almost parallel to the $x-$axis:  their estimates of $x_E$  were widely spread, ranging from 2000~m to 4200~m with a 
median around 3100~m.
Conversely, witnesses
whose line of sight was almost parallel to the $y-$axis
were better able to estimate $x_E$. Compiling their accounts,
$x_E\approx 2900 \pm 400$~m.

(iii) The third determination uses rough estimates of the aircraft speed and of the fall duration $D=t_I-t_E$.
According to witness ${\cal F}$,
$D$ was shorter than 15 seconds. 
This is compatible with  a lower bound for $D$ of 6.2~s (Section~\ref{sec:free_fall}).
Assuming arbitrarily that $D=12 \pm 3$~s,
 using our rough estimate of ${\dot{x}}_I=
 -72.7 \pm 10$~m/s (Section~\ref{sec:speed_at_impact}), 
 and since  the aircraft inertia implied that ${\dot{x}}$ only changed by a few percents between $E$ and $I$, from (Eq.~(\ref{eq:speed_profile}))
 we obtain 
$x_E \approx x_I  + D \;  {\dot{x}}_I
= 3030 \pm 300$~m.

The 
three determinations of $x_E$ together yield
$x_E=\modifxE{2970} \pm \modifxE{345}$~m.
This value is compatible with the lower bound found using the free fall model, $x_E > 2570$~m (Section~\ref{sec:free_fall}). 
As discussed in Section~\ref{sec:sensitivity}, this value affects several other quantities calculated below (Table~\ref{table:results}).

As a by-product, this value of $x_E$ enables to estimate the following  quantities.
First, $z_E-z_O = 
x_E \; \tan \gamma
 = \modifxE{154}
\pm \modifxE{31}$~m.
Hence the aircraft had almost completed its descent. In terms of absolute altitude, 
$z_E = z_O +  \modifxE{154} =  \modifxE{1639} \pm  \modifxE{31}$~m; 
laterally, $y_E=0$,  with {an uncertainty of} $8.7\; 10^{-3} {x_E} = \modifxE{26}$~m.
Eq.~(\ref{eq:speed_profile}) yields the aircraft speed at interception, ${\dot{x}}_E=
-77.1$~m/s,
with {an uncertainty} that we roughly estimate to be $\pm 10$~m/s.
\label{sec:pos_and_speed_interception}
We also retain  a vertical speed ${\dot{z}}_E 
= {\dot{x}}_E \; \tan \gamma
 = -4.0\pm 0.5$~m/s. The determination of  ${\dot{x}}_E$ feeds back on  the estimate of the fall duration: we refine it by using the distance between $I$ and $E$ and assume a constant deceleration between 
 $I$ and $E$. It yields
$D=11  \pm 2.5$~s,  compatible both with the lower bound of 6.2~s (Section~\ref{sec:free_fall}) and the upper bound of 15~s of witness account ${\cal F}$.

This fall duration, assuming the vertical acceleration is constant (but lower that than of gravity), and thus assuming a parabolic vertical trajectory, would yield a vertical acceleration $\modifxE{3.06} \pm 1$~m/s$^2$.
In practice, this value is an average over the trajectory, since during the fall the vertical acceleration may have varied  due to several factors.
Neglecting these variations, we find 
a vertical speed at impact of $\modifxE{-37.7}$~m/s (confidence interval \modifxE{29.3} to \modifxE{43.7}~m/s)
and a vertical angle at impact of $\modifxE{26.1}^\circ$ (confidence interval \modifxE{21.3} to \modifxE{30}$^\circ$);
the lateral angle at impact being $\tan {\beta} = 13.7^\circ$, the lateral speed ${\dot{y}}_I$ was of order of ${\dot{x}}_E \;  \sin {\beta}= -18.3 \pm 2.3$~m/s. 
\label{sec:impact_angle_orientation}

Based on simple approximations on the trajectory during the fall and on its duration, we thus find a consistent picture. We also confirm that the free fall figures (Section~\ref{sec:free_fall}) were upper bounds for the vertical acceleration and vertical impact angle, and  lower bounds for $x_E$, $z_E$, $-{\dot{x}}_E$ and $D=t_I-t_E$. 
{We again refer the reader to Table~\ref{table:results} to keep track of these results.}

\section{Missile}
\label{sec:missile}

In ocular and/or auditive accounts~\cite{mutsinzi} the number of missile launches ranged from 1 to 3 missiles (median and maximum at 2) 
 \label{sec:missile_number}
and the time interval between them from a fraction of second to many seconds (median at $\sim 2$~s).
No witness has reported that  missiles could have been launched simultaneously from different locations. 
There is a consensus that only one missile successfully intercepted the aircraft. 
In what follows, we focus on this successful missile.
We now examine  witness accounts to extract information on the  missile   launch site (Section~\ref{sec:missile_launch_site}), trajectories (Section~\ref{sec:trajectory_orientation}) and type (Section~\ref{sec:missile_type}).

\begin{figure}[t]
\centering
\showfigures{\includegraphics[width=0.75\textwidth]{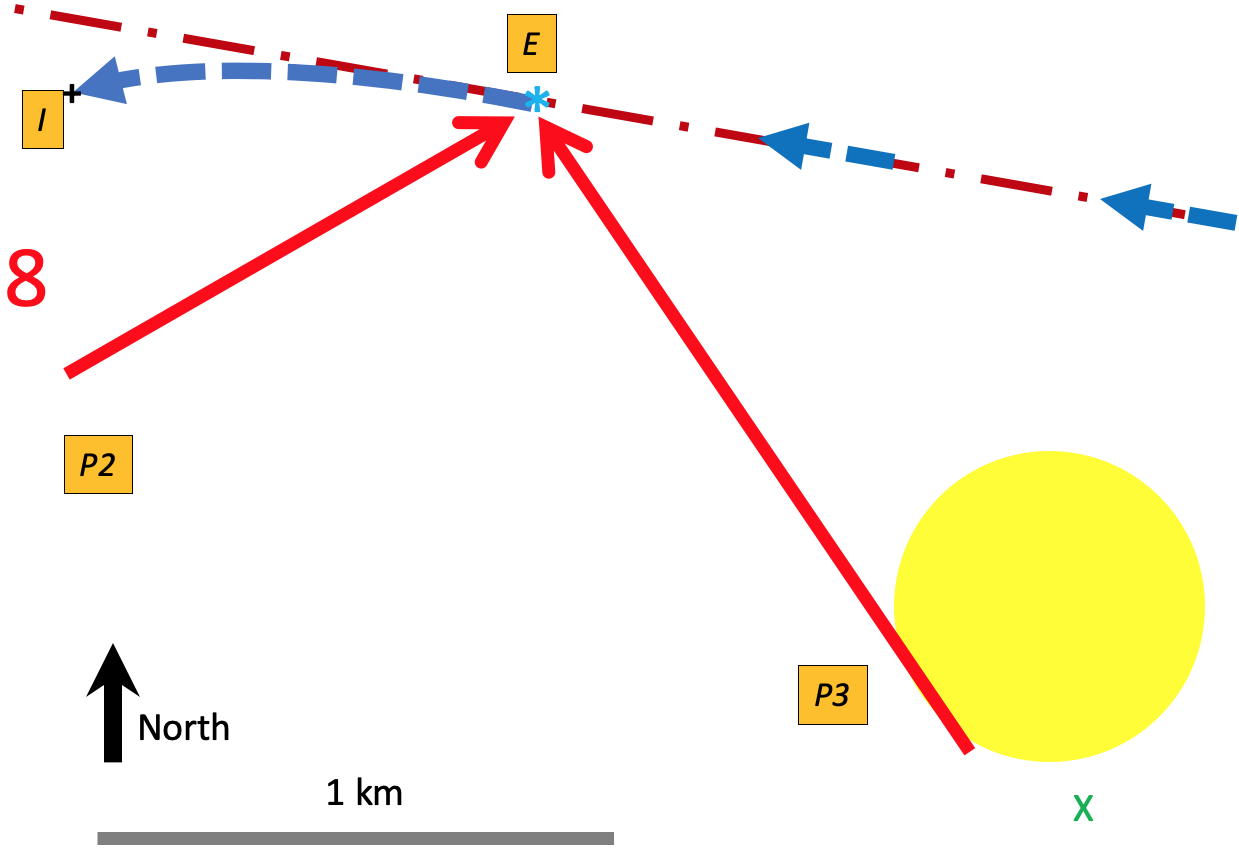}}
\caption{{\bf Possible missile launch sites and trajectories.}
Scale bar: 1 km.
Brown dash-dotted line indicates  the standard aircraft trajectory along the $Ox$ axis, 
and the thick blue dashed arrows, pointing left, indicates the direction of the aircraft's approach; 
the  position reported by the crew and the 
point at which an aircraft should reach the ``glide" (prescribed trajectory) are not shown and lie to the right of this image. 
$E$ marks the aircraft interception by the missile (blue star), as estimated backwards from the impact data (Section~\ref{sec:longitudinal_fall}); $I$ marks the aircraft impact on the ground (black cross).
Thick red solid arrows mark possible missile paths until $E$. In this top view, they are almost straight (Section~\ref{sec:trajectory_orientation}). 
They are drawn from starting from to two possible launch locations according to witness accounts: either South-West of $E$ (left), derived from Pair 2 (see their accounts in Fig.~\ref{fig:interception_position}) and number ``8" 
 (Fig.~\ref{fig:triangulation}); or
South-East of $E$ (right), derived from Pair 3 (Fig.~\ref{fig:interception_position}), numbers ``4" to ``7" 
 (Fig.~\ref{fig:triangulation}) whose uncertainty range is represented here by a yellow circle, and intersection  (green cross) of accounts of missile launcher location (Section~\ref{sec:supp_missile_launchers}).
}
\label{fig:synthesis}
\end{figure}

\subsection{Missile   launch site}
\label{sec:missile_launch_site}

Several witness accounts could contribute to determine the site $L$ from which the missile was launched. However, they can be subjective, imprecise, or contradictory due to memory distortions, and corresponding uncertainties have to be taken into account. We have examined a few possible methods to interpret the available information: distance from witness to launch site (``trilateration", Section~\ref{sec:supp_trilateration}),  perceived direction of the launch site  (``triangulation", Section~\ref{sec:supp_triangulation_directions}).
Depending on the method to select the witness accounts and to discard outliers, we find two main locations for $L$ (Fig.~\ref{fig:synthesis}). 
One
is at the South-West of the aircraft interception: 
 $x  = 2092$~m, $y = -480$~m, with a  r.m.s. distance $\sqrt{\bar{d^2}} = 328$~m.
The other is at the South-East of the aircraft interception: $x = 4000 \pm 100$~m, $y = - 1130 \pm 30$~m, 
also compatible with the 
alleged finding of missile launchers on the ground  (Section~\ref{sec:launch_pos_value}). We choose the latter as an example for the calculations performed in Sections~\ref{sec:trajectory_orientation},
 \ref{sec:missile_type} and in
Table~\ref{table:results}.

\begin{figure}[t]
\centering
(a) Top view \hfill \hfill (b) Side view \hfill \hfill (c) Foreshortened view  \hfill ~ \\
 \hfill ~ \\
\showfigures{\includegraphics[height=0.275\textwidth]{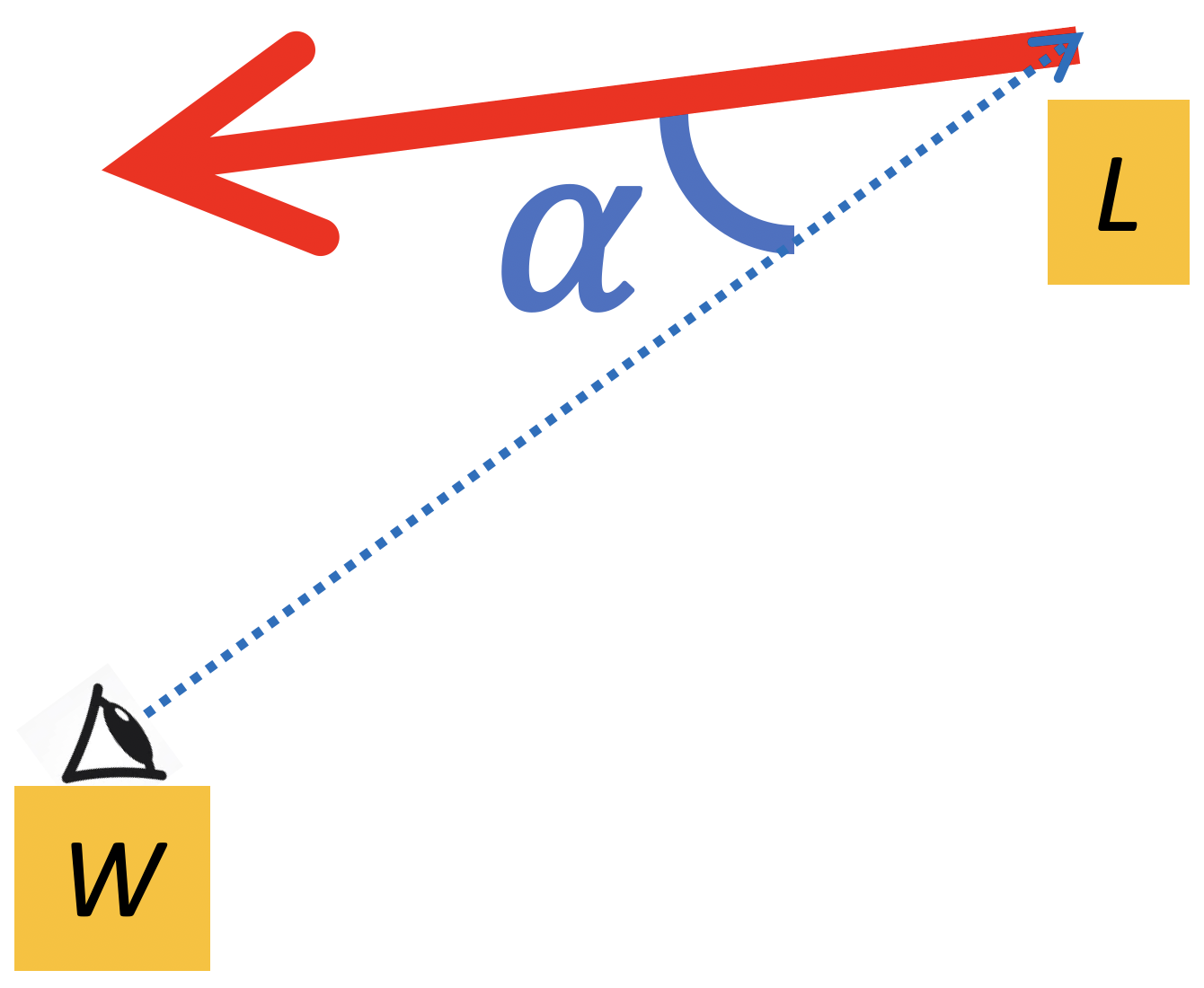}} \hfill 
\showfigures{\includegraphics[height=0.275\textwidth]{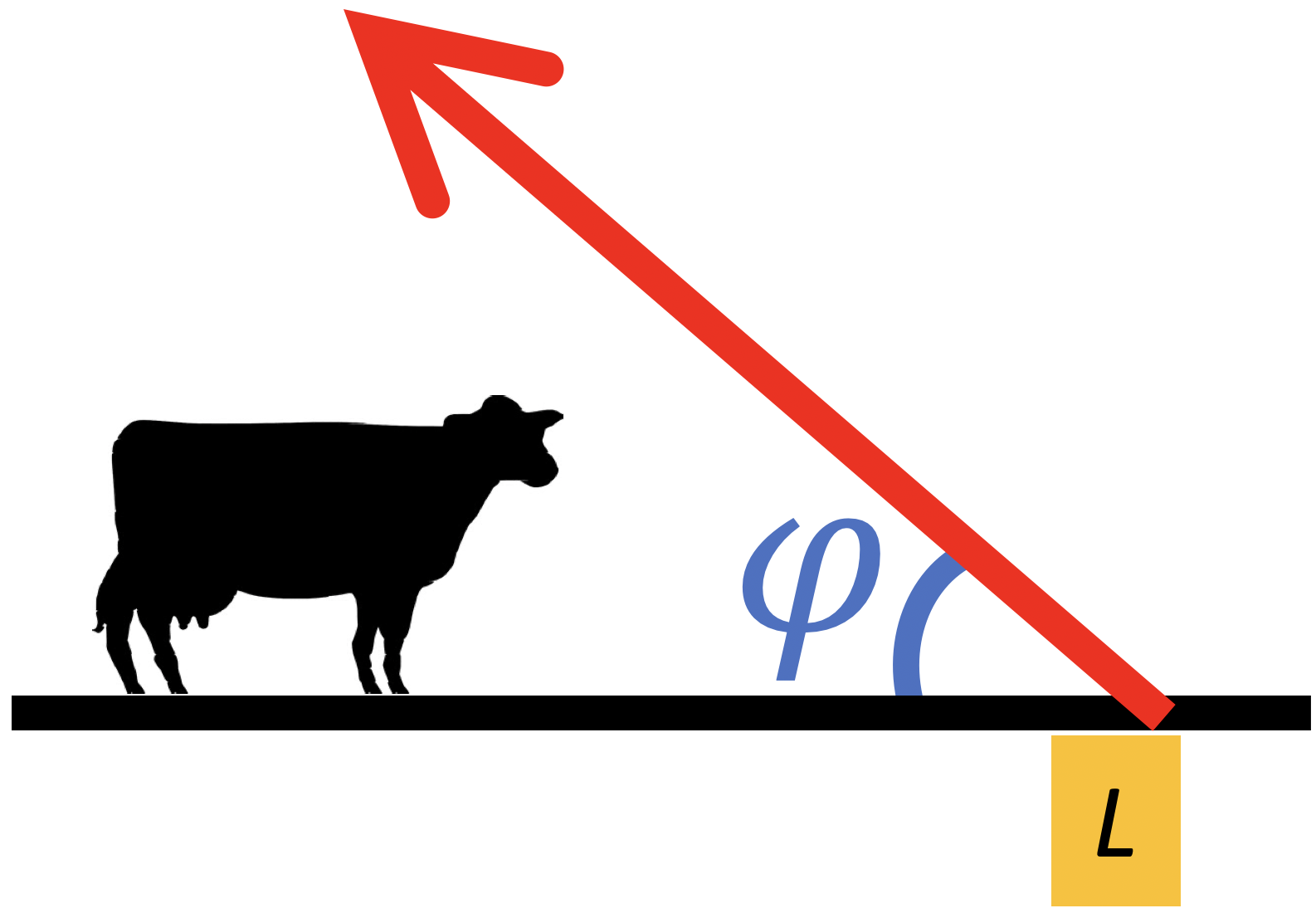}} \hfill 
\showfigures{\includegraphics[height=0.275\textwidth]{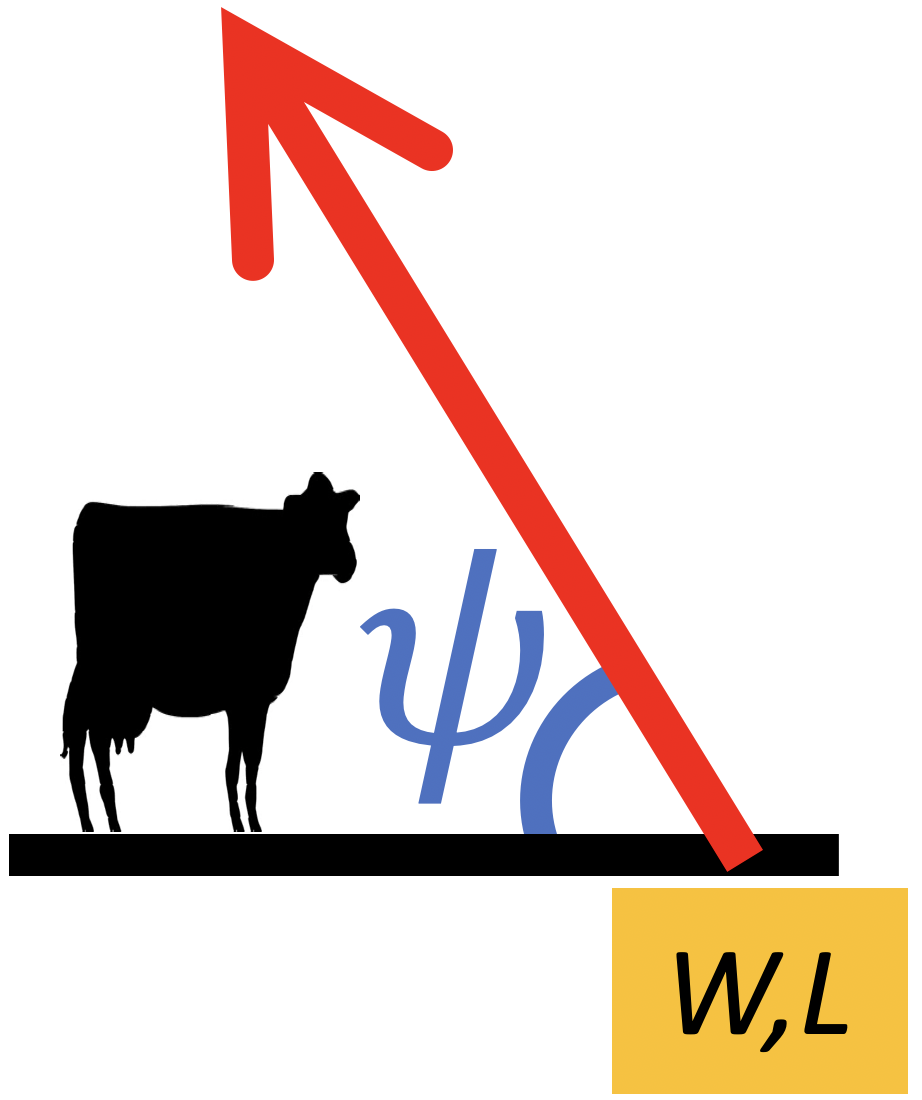}}\\
\caption{{\bf Schematic of angle between missile trajectory and horizontal plane.}
(a)  
Overhead view of the missile path. $W$: witness position; dashed blue arrow: the witness' line of  view $WL$;  red arrow: the missile trajectory's vertical plane;  $\alpha$: angle between $WL$ and the missile trajectory projected on the horizontal $Oxy$ plane.
(b) View of the missile path from the side.
The actual launch angle $\varphi$  with respect to the horizontal plane $Oxy$ is seen in the vertical plane containing the missile trajectory. 
Thick black line: ground, i.e. horizontal $Oxy$ plane; $L$: launch site; red arrow: initial tangent to the trajectory.
(c) Foreshortened view from the witness' location. The apparent missile trajectory angle $\psi$ with respect to the horizontal  $Oxy$ plane is perceived by the witness along the witness' line of sight. In such projection, $W$ and $L$ coincide, and horizontal components of panel (b) elements are foreshortened by a factor $\sin \alpha $ (equal to 0.5 in this example). 
}
\label{fig:apparent_angle}
\end{figure}

\begin{table}[t]
\centering
\caption{ {\bf Apparent angle between missile trajectory and horizontal plane.}
For definitions and notations see Fig.~\ref{fig:apparent_angle}. Values  of $\psi$ as indicated by four witnesses;
 confidence interval (minimum; mean; maximum) of $\alpha$ values estimated  using the witness location $W$, the launch location $L$ and the trajectory orientation 
 confidence interval  (\modifxE{119.8}$^\circ$; \modifxE{132.3}$^\circ$; \modifxE{141.9}$^\circ$)
 with respect to the $x-$axis; 
 estimates of $\varphi$ and confidence interval based on $\psi$, $\alpha$ and Eq.~(\ref{eq:phi_alpha_psi}).}
\begin{ruledtabular}
\begin{tabular}{l c c c}
Witness

 & $\psi$ & $\alpha$  &  $\varphi$  \\
\hline%
\hline
${\cal E}$ 
  
 &  60$^\circ$  &  (\modifxE{123.8}$^\circ$; \modifxE{136.3}$^\circ$; \modifxE{145.9}$^\circ$) &  (\modifxE{44.2}$^\circ$; \modifxE{50.1}$^\circ$; \modifxE{55.2}$^\circ$) \\
${\cal J}$
 
  &  43$^\circ$ &  (\modifxE{126.8}$^\circ$; \modifxE{139.3}$^\circ$; \modifxE{148.9}$^\circ$)  &  (\modifxE{25.7}$^\circ$;  \modifxE{31.3}$^\circ$; \modifxE{36.7}$^\circ$)  \\
${\cal L}$ 

 & 70$^\circ$  &  (\modifxE{180.8}$^\circ$;  \modifxE{193.3}$^\circ$; \modifxE{202.9}$^\circ$)  & (\modifxE{02.2}$^\circ$; \modifxE{32.3}$^\circ$; \modifxE{46.9}$^\circ$)   \\
${\cal M}$
  
   & 45$^\circ$  &   (\modifxE{129.8}$^\circ$; \modifxE{142.3}$^\circ$; \modifxE{151.9}$^\circ$)  &  (\modifxE{25.2}$^\circ$;  \modifxE{31.4}$^\circ$;  \modifxE{37.5}$^\circ$)   \\

\end{tabular}
\end{ruledtabular}
\label{table:apparent_angle}
\end{table}

\subsection{Missile trajectory}
 \label{sec:trajectory_orientation}

 We now examine the missile trajectory (length, orientation and curvature) which, as we discuss below  (Section~\ref{sec:missile_type}), {may} have an influence on the  determination of the missile type. 
We discuss  the trajectory 
in the horizontal plane, as would have been viewed from the sky, 
and in its vertical plane, as viewed by the witnesses.

From  the estimates of the missile launch site $L$ (Section~\ref{sec:launch_pos_value}) and of the  interception location $E$ (Section~\ref{sec:pos_and_speed_interception}), we infer that the trajectory endpoint coordinate differences $(x_E-x_L, y_E-y_L, z_E-z_L)$ 
were of order of
 $(\modifxE{-1030},+1130,\modifxE{+305}) \pm (\modifxE{359},40,\modifxE{31})$~m. 
Hence the missile trajectory had a length ${\ell_m}$ of order of $\modifxE{1559} \pm \modifxE{362}$~m,
and
an angle within the horizontal plane $Oxy$  with respect to axis $Ox$  of $\modifxE{+132} \pm \modifxE{11}^\circ$ 
	(red arrows on Fig.~\ref{fig:synthesis}), and an angle (averaged over the whole trajectory)  with respect to the horizontal plane $Oxy$ of  $12 \pm \modifxE{4}^\circ$.

Four witnesses reported a value for the launch angle, $\psi$, with respect to the horizontal plane $Oxy$
 (Table~\ref{table:apparent_angle}).
The actual launch angle $\varphi$  with respect to the horizontal plane $Oxy$ can be deduced from the one perceived by the witness, $\psi$, using the relation (Fig.~\ref{fig:apparent_angle}):
\begin{equation}
\tan \varphi = \tan \psi \; \vert \sin \alpha \vert 
\label{eq:phi_alpha_psi}
\end{equation}

For each of these witnesses, we determined $\alpha$ assuming a launch site  $L$ located  in the South-East of the interception, 
and accordingly the missile trajectory orientation  
$\modifxE{+132} \pm \modifxE{11}^\circ$
with respect to the $x-$axis. 
The average of the four   values is  
$\varphi = \modifxE{36.3}^\circ$ with a confidence interval $\modifxE{20.4}^\circ$  to $\modifxE{45}^\circ$.
With this location for $L$, we find no inconsistency between the sign of $ \sin \alpha$ and the witness accounts reporting that they saw the missile ascending  
 ``from right to left" (e.g. {\small \it ${\cal E}$})
or
``from left to right" (e.g. {\small \it ${\cal L}$}). 
Note that for  the alternative launch location, 
 we  find incompatibilities both in the values of $\varphi$ and in the signs of $\sin \alpha$ (data not shown).

Some witnesses   saw the trajectory  curved downward in the vertical plane, with a more or less pronounced curvature  depending on their position,
This explains the difference between the initial launch angle, $ \varphi = 36\pm 10^\circ$, and the average trajectory angle,   $ \modifxE{12} \pm  \modifxE{4}^\circ$ (see above).
To fix the ideas, if the curvature was constant, the difference between the curve length (arc) and the distance  $\modifxE{1559} \pm \modifxE{362}$~m
between the endpoints (chord) would be less than 1~m.

The  curvature in the horizontal plane, corresponding to the evolution of $\alpha$ in time,  was not mentioned in the witness accounts and has to be inferred from other  information.
The missile was  homing to the aircraft  (Section~\ref{sec:missile_type}) and thus may have changed $\alpha$ to adapt to the change in aircraft position during the time interval between launch and interception.  The average speed $c_M$ of the missile is  of order of ten times the aircraft speed at interception ${\dot{x}}_E$. Thus the change in missile direction $\alpha$ was at most 20$^\circ$ over 1.5~km, corresponding to a  curvature radius at least of order of 5~km:
the  trajectory has barely been curved in the horizontal plane.

\subsection{Missile type}
 \label{sec:missile_type}
 
We have also looked for quantitative arguments to discriminate between surface-to-air weapon types. 
In 1994, hundreds of types were marketed (Ref.~\cite{oosterlinck} pp.~102-174),
that fall into these three classes:  non propelled (e.g. machine gun bullet); self-propelled but non-homing (rocket); or self-propelled and 
homing (missile), guided by the infrared emission the aircraft's hot engine exhaust gases.
Weapons from all three classes can at least travel over a distance of 2000~m at 300~m/s and rise to a height of 1500~m, meaning that they all had the capability of intercepting an
 aircraft flying  \modifxE{1559}~m away at a speed of 77.1~m/s and at 
an altitude of 
\modifxE{154}~m.
Projectile trajectories were described as luminous,
 which could be due either  to a missile's motor exhaust (Ref.~\cite{cranfield} p. 18-19)
 or to tracer ammunitions. 
The trajectory itself provides an information.
If the  projectile had been purely ballistic, gravity
would 
have affected  its speed   by a few tens of m/s and its altitude by a few tens of meters, 
so that its trajectory would only have been marginally affected.
By contrast, the trajectory  we have inferred had an initial launch angle of $36\pm 10^\circ$  with an average value of $12 \pm 3^\circ$ over the entire trajectory (Section~\ref{sec:trajectory_orientation}). 
It is thus curved, which is indicative of guidance.
This was confirmed by witnesses 
${\cal H}$,
${\cal O}$,
and ${\cal N}$
who saw the trajectories bend and head towards the plane.

In 1994,  the most widespread missile types were the Stinger produced in the USA,  the Strela/Igla/SA/SAM  produced in the USSR/Russia, the Mistral produced in France, and copies  thereof developed in other countries (Ref.~\cite{oosterlinck} p.~108-109).
For simplicity, we compare only the specifications of these three missile types.
Many of the launch characteristics are compatible with all three types of missile.
It was usual for these  missiles to be launched at a large angle from the horizontal before their trajectories bended towards the horizontal.
For instance, the Mistral has a nominal launch angle that ranges up to 67$^\circ$~\cite{simbad}.  All missiles would had to be launched far enough from their target, since the guidance was operational only beyond a minimal distance: 500~m for Mistral~\cite{simbad}, 1000~m for SAM (Ref.~\cite{oosterlinck} p. 319).
By contrast, a non-propelled weapon, which would have performed a direct shot, could have been launched from a place much closer to the aircraft trajectory.

In principle, the luminous trace  could help discriminate the missile type (Ref.~\cite{oosterlinck} p. 321).
Several witnesses
(e.g.  ${\cal J}$,
${\cal N}$)
 indicated that they had seen the  luminous trace until it intercepted the aircraft.
A missile specialist suggested that the trajectories for the SAM or Stinger missiles are  visible over their whole length, while the trajectory of a Mistral missile is visible only during the first 2~s~\cite{ancel_nepassubir}. 
Conversely, UK experts estimated that the SAM trajectory is visible only over the first 2~s (Ref.~\cite{cranfield} p. 18-19).

We will now examine these claims. The publicly available information on the missiles includes either the duration $t_p$ of the propulsion phase, its acceleration $a_p$ or its final speed $v_p$.
During this propulsion phase, the engine flames make the trajectory visible. In principle, we could compare the length (or duration) of this phase with the  determined value for the length (or the duration) of the trajectory, then determine how much further (and longer) the missile must have flied during the non-propulsion period. This could be an interesting but complicated problem involving all details of the successive missile propulsion phases, the variation of its acceleration with time, or the changes in speed during the non-propulsion phase. 
For simplicity, we approximate the missile trajectory  with two phases: a phase at roughly constant acceleration during which the engine flames make the trajectory visible and a less accelerated, less  visible  phase at roughly constant speed  (Section~\ref{sec:supp_miss_duration}). Results based on these assumptions are reported in  Table~\ref{table:missile_trajectory_duration}.
For all three missile types, the total trajectory duration
is  slightly larger than that of the  visible phase.
In the case of a Stinger missile, the missile could even not have reached the less-visible phase,  so its luminous trace would have appeared to intercept the aircraft.
Conversely, for a Mistral missile the less-visible phase lasts at least \modifxE{0.1}~s (\modifxE{83}~m), and for a SAM missile  \modifxE{0.74}~s  (\modifxE{511}~m).

\section{Discussion}
\label{sec:conclusion}

\label{sec:sensitivity}

This study illustrates how physics courses can apply to a real-life situation, as well as the importance and limits of scientific expertise during a judiciary process.
Each section  provides at least one idea of exercise for students. In addition, the impact for teaching is far-reaching, for the following reasons.
The methods developed are in themselves a subject to teach. The students can be encouraged to provide data, determine which ones are important to discriminate between alternative models, determine the fields of physics which can feed the reasoning, and prevent circular reasoning while inferring information from whatever existing data. The uncertainty discussion can train the students' critical mind, as well as teach methods (e.g. inference, outlier trimming, optimisation) routinely used in several fields of physics research (e.g. in multi-messenger astrophysics, in complex systems studies, or in particle physics). 
They can learn how the uncertainties regarding data propagate to determine results uncertainties.

We hade to make methodological choices and assumptions, and they affect the results.
Our  methodological choices   include:
discarding outliers;
choosing minimization based on witness accounts;
or neglecting variants of each missile type.
Our assumptions  include:
assuming the aircraft has followed the standard instrument approach procedure;
neglecting variations of aircraft acceleration; 
simplifying  the missile trajectory phases;
neglecting aircraft's velocity vector changes when it was intercepted by the missile;
or ignoring 
evasive maneuver
 by the pilot.

We keep a trace of justifications and consequences of our choices. We chose an open approach, i.e. we avoid any a priori hypothesis on the launch site: the list of accounts is compiled, and the launch site is inferred by minimizing the global disagreement with the various witness accounts, possibly weighted. We try to make our choices as explicit as possible, to support them with arguments, and to minimize their effects, e.g. by keeping a maximum of data and discarding only obvious outliers using standard procedures. The list of accounts we use is subject to biases, thus we state it explicitly.
We perform consistency checks. 
Several values that we calculate (Table~\ref{table:results}) are independent from each other, i.e. a mistake or a rounding of one of these values does not modify the other values. Other values are inter-dependent, for instance a change in $x_E$ propagates and creates a spate of changes, possibly with non-linear effects. As a test of sensitivity and traceability, we round $x_E= 2970 \pm 345$~m as $3000 \pm 300$~m. Table~\ref{table:results} shows  in tiny italics the corresponding results.

\section*{Supplementary Material}

Please click on this link to access the supplementary material, which includes additional details and possible studies. 
Print readers can see the supplementary material at [DOI to be inserted by AIPP].

\begin{acknowledgments}

We thank 
Jacques Morel and Aymeric Givord
for maintaining and making  the source document database (\url{http://francegenocidetutsi.org}) accessible, 
Guillaume Ancel 
for technical explanations regarding missiles,
Timothy Van Renterghem and Jos van Oijen for pointing out mistakes in the initial version of the manuscript,
colleagues from the
MSC 
laboratory for stimulating discussions and especially 
Caroline Derec 
for critical reading of the manuscript. 
For the purpose of Open Access, a CC-BY 4.0 public copyright licence \url{<https://creativecommons. org/licenses/by/4.0>} has been applied by the authors to the present document and will be applied to all subsequent versions up to the Author Accepted Manuscript arising from this submission. 

\end{acknowledgments}

\section*{Conflicts of interest}

The authors have no conflicts to disclose.

\section*{Author contributions}

All authors  contributed equally to this work. F.G. wrote the initial draft, which was edited by S.M.P.

\clearpage

\beginsupplement
 
\begin{center}
    \LARGE\textbf{Supplementary Material}\\[1em]
  {\it \large ``Inference uncertainty about an aircraft crash" }\\[1em]
    \large{Fran\c{c}ois Graner, Stefano Matthias Panebianco}
\end{center}

\vspace{2em}

This Supplementary Material lists additional details and possible studies.

\section{Aeronautic units and data}
\label{sec:supp_aero_units}

A foot is 0.3048 m, 
a mile is 1852~m, a knot is 0.514 m/s, a dot is $0.5^\circ$.
Aeronautic maps use the magnetic North; the difference from geographic North (the ``declination") was
3$^\circ$~West for the 1999 map (Ref.~\cite{oosterlinck} p.~185), 0$^\circ$ for the 2006 one (Ref.~\cite{cranfield} p.~106), 
1$^\circ$~East for the 2024 one~\cite{rcaa}, so that the direction of the airport track (called ``runway 28")  was marked as $77^\circ$, $80^\circ$  or $81^\circ$, respectively, on these maps.
We  consistently use the geographic North, so that the runway orientation is  $80^\circ$ whatever the year.
Kigali international Airport was at that time called Gr\'egoire Kayibanda international airport, and on later maps appears also as
 IATA code KGL, ICAO code HRYR, or  Kanombe International Airport.

\section{Free fall hypothesis and its inconsistencies}
\label{sec:free_fall}

\begin{figure}[t]
\centering
\showfigures{\includegraphics[width=1\textwidth]{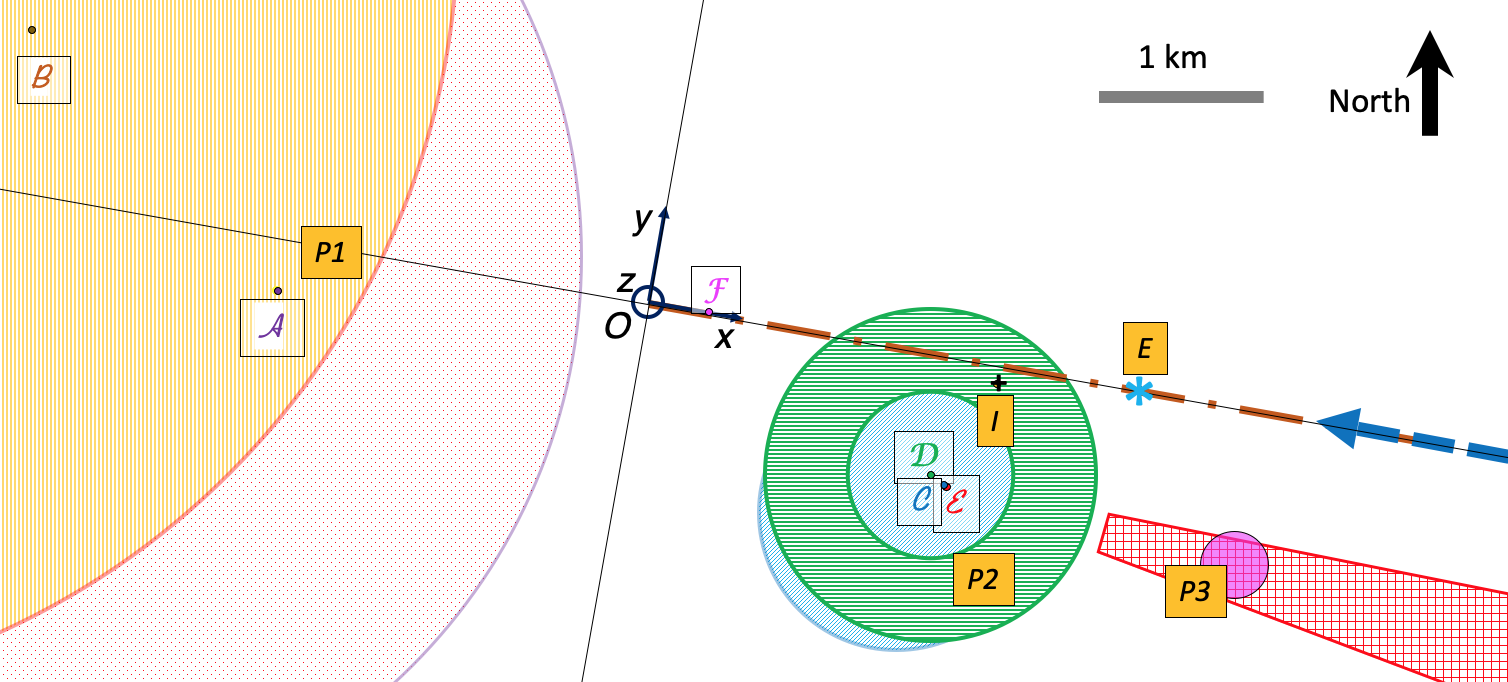}}
\caption{{\bf Distance information on  missile launch position.}
Same caption as Fig.~\ref{fig:synthesis}.
Letters and colored shapes indicate witness locations and information gained from their testimony, explained in Sections~\ref{sec:longitudinal_fall}, \ref{sec:free_fall}, \ref{sec:supp_trilateration}:
${\cal A}$, purple dotted circle;
${\cal B}$, light brown  vertically hatched circle, 
${\cal C}$, blue diagonally hatched circle;
${\cal D}$, green  horizontally hatched ring;
${\cal E}$, red cross hatched sector;
${\cal F}$, violet circle.
}
\label{fig:trilateration}
\label{fig:interception_position}
\end{figure}

To find the position $x_E$ at which the missile intercepted the aircraft, we 
have to connect the aircraft's trajectory after the interception with the planned trajectory
(Fig.~\ref{fig:schema_trajectories}a).
As a first approximation, let us  examine the unlikely but simplest possible trajectory for the aircraft after its interception, that of a free fall studied by French investigators (Ref.~\cite{oosterlinck} p. 189-193).
If the plane had  stalled, its trajectory projected on the vertical $Oxz$ plane would have been a ballistic parabola  (Fig.~\ref{fig:schema_trajectories}b) of vertical acceleration ${\ddot{z}} = -g=-9.81$~m/s$^2$; and  on the horizontal $Oxy$ plane it would have been a straight line (Fig.~\ref{fig:schema_trajectories}a), which would require a discontinuous change in direction at the time of interception. 
Extrapolating the fragment dispersion angle (Section~\ref{sec:orientation_fragment}), the interception point $E$ would then be situated upstream of the ground impact at
$x_E-x_I = {(y_E-} y_I  {)} / \tan {\beta}$.
Here
$x_I =  2 160  \pm 20$~m,
 $y_I = -100 \pm 20$~m,
$ {\beta} = 13.7 \pm 4^\circ$,
and $y_E=0 \pm 8.7\; 10^{-3}\; x_E$.
 Altogether,  this yields $x_E-x_I = 410$~m,
and assuming that the  confidence intervals are independent 
we obtained $x_E=2570 \pm 160$~m,
 $y_E=0\pm22$~m (Fig.~\ref{fig:interception_position}).
This  determines the fall duration $D=t_I-t_E$ and the speeds ${\dot{x}}_E$, ${\dot{x}}_I$. 
Using  Eq.~(\ref{eq:glide_slope_b}),
 one  obtains that at interception, $z_E = z_O + x_E \; \tan \gamma
 = 1620$~m and 
${\dot{z}}_E =  {\dot{x}}_E \; \tan \gamma$.
Eq.~(\ref{eq:speed_profile})  yields that 
${\dot{x}}_E=
 -74.9$~m/s and ${\dot{z}}_E=-3.9$~m/s.
The vertical trajectory
combined with the impact position $I$  yields a quadratic equation for $D$, whose positive solution is $D=6.2$~s. 
The vertical speed at impact then is ${\dot{z}}_I=-64.7$~m/s.

 This free fall scenario seems unlikely for several reasons.
First, while this scenario is built
such that the lateral deviation $\beta$ is correct,
it relies on a very specific and unlikely combination of factors,
see Sections~\ref{sec:momentum_transfer}, ~\ref{sec:lateral_fall}.
Producing a deviation of 13.7$^\circ$ when ${\dot{x}}_E=
-77.1$~m/s
 (Section~\ref{sec:longitudinal_fall}), the discontinuous change in velocity 
caused by the interception would 
be 18.2~m/s,  which is much larger than the above estimates (Section~\ref{sec:aircraft_mass_before}). Moreover,
 it  would require a change in velocity which,  
  by coincidence, has been  such that  the aircraft orientation changed by 13.7$^\circ$ during the fall. In addition, in order to explain the sign  of both the center-of-mass and angular velocity changes,  the missile or its explosion should have affected the aircraft on its right (northern) side,  near the nose rather than the tail.

Second, free fall requires stalling. In reality,
between interception and impact,
the air flow around the intact right wing  probably remained laminar for some time, changing progressively rather than discontinuously. 
The aircraft did probably  lose  lift on the left wing due to  damages, but kept enough lift to partially compensate its weight for a significant time.
It means while gliding the aircraft could probably have avoided stalling  (Fig.~\ref{fig:schema_trajectories}b).
In fact, pictures of the aircraft debris (Ref.~\cite{oosterlinck} p.~95) show that, as expected  before landing and for a speed lower than 98 m/s~\cite{manuel}
the pilot had already extended the  high lift devices (so-called ``flaps"). 
In this landing configuration, the speed at which  a Falcon 50  stalls would typically have been of order 42~m/s:
the actual speed was well above this limit.

Third,
the value of vertical speed according to the free fall model, ${\dot{z}}_I=-64.7$~m/s, would imply a vertical impact angle 
$\arctan({\dot{z}}_I/{\dot{x}}_I)
=41.7^\circ$,
which cannot explain the observed bouncing on the ground.
Indeed, one French military investigator observed soon after the crash 
that no debris was impaled into the ground, that there was no crater around the debris corresponding to the back part of the plane, and that the front part had continued its longitudinal sliding while creating a furrow; he inferred that only a small part of the plane had been damaged by the missile, that the plane had slowed down progressively, and that it was roughly horizontal at the time of impact (Ref.~\cite{bruguiere}  8567, p. 12).
Belgian investigators noted that, although the ground was soft, the impact crater due to the front part was shallow:
they estimated 
the aircraft impact angle being at most 20$^\circ$ to the
horizontal  (Ref.~\cite{belgian_investigators} p.~3). 
UK experts agreed with this estimate and added that aircraft debris would have been expected to be
found in the crater if the aircraft had adopted a more vertical descent
(Ref.~\cite{cranfield} p.~8).

All these remarks make this free fall scenario unlikely. We thus consider the free fall estimates only as upper bounds for the vertical acceleration and vertical impact angle, and as lower bounds for $x_E$, $z_E$, $-{\dot{x}}_E$ and $D=t_I-t_E$.

\section{Trilateration, using distances}
\label{sec:supp_trilateration}

Some witness accounts can be translated into  information  regarding distances. 
In principle, they could be combined to determine  the missile launch site $L$  via a method called ``trilateration". 
We observe that they belong to three categories.

\label{sec:triangulation_sound_amplitude}
First, sound level could provide information on distance. The acoustics expert had to determine how the sound level decreases with distance (Ref.~\cite{oosterlinck} p.~224). His conclusion was that the missile launch sound level was perceived as 160~dB near the launch site, 133~dB at 80~m, 125~dB at 200~m, 104~dB at 2100~m (Ref.~\cite{acoustic} p.~18).
However, since  witness accounts are not quantitative, the direct comparison is difficult. 
What can be interesting for students is to understand the theoretical equation describing how the sound level decreases with distance in empty space.
 If we note $L_p$ the acoustic power emitted by the missile launch, and $L_w$ the acoustic pressure at a distance $d$ from the launch site,
 the acoustic power is distributed over a sphere of surface area  $S = 4\pi d^2$. The dB requires its decimal logarithm.
 Hence the term $10 \log_{10}(1/S) = - 20 \log_{10}(d) - 10 \log_{10}(4\pi)$, and finally  $Lp \approx L_w  - 20 \log_{10}(d) - 11$.
 This corrects an error of sign in the expert's report (11 instead of $-11$) who overestimates sound levels by 22~dB.
In practice, sound levels variations depend on multiple other  factors taking into account the ground, its relief and the various obstacles.

\label{sec:triangulation_sound_speed}

Second, three  
witnesses
${\cal A}$,
${\cal B}$,
${\cal C}$,
mentioned a specific time sequence: these witnesses have heard the noise coming from the launch before they saw the missile trajectories (or at least part of them).
In principle, this could indirectly provide an information, namely an inequality between the distance $\ell_{LW}$  between launch site $L$ and witness location $W$, and the distance $\ell_{LE}$ between launch site $L$ and aircraft interception location $E$: $\ell_{LW} / c_s  <  \ell_{LE} / c_M$, where $c_M$ is the missile average speed.
Equivalently, this inequality can be rewritten as: 
\begin{equation}
 \frac{\ell_{LE}}{\ell_{LW}} > \mathrm{Ma}
\label{eq:inequality_elllipse}
\end{equation}
where $\mathrm{Ma}= c_M / c_s$ is the missile Mach number. Its order of magnitude is 2 (Section~\ref{sec:missile_type}) and we retain this value to fix the ideas. 
Then the  set of  points $L$ obeying Eq.~(\ref{eq:inequality_elllipse}) is a circle (Fig.~\ref{fig:trilateration}) of  radius $2\ell_{WE}/3$,
and  whose center is located at a distance  $\ell_{WE}/3$ from $W$.
In practice, these three accounts are not mutually consistent (Fig.~\ref{fig:trilateration}).

Third,  three witnesses provided distance estimates from their location to the launch site without explaining explicitly how they had obtained them, so that the information provided should be taken with care.
${\cal D}$ 
reported having 
had much experience in hearing shootings and suggested a distance between 500 and 1000~m.
${\cal E}$
saw the launch at 1000 to 5000~m distance and reported its direction.
${\cal F}$
explicitly stated that the launch took place near
a factory (violet circle on Fig.~\ref{fig:trilateration}).
At that time, the factory was  a 80~m~$\times$~40~m rectangle  (Ref.~\cite{bruguiere}, 6017) centered at $x=3660$~m and $y=-870$~m (see below, Fig.~\ref{fig:synthesis}).
Given that its size was less than 100~m and its distance to the closest other landmarks was more than 300~m, we arbitrarily attribute to this account a confidence interval of 200~m. 
Again, these three accounts are not mutually consistent (Fig.~\ref{fig:trilateration}).

In summary, three pairs of distance accounts ({combined with direction accounts whenever available}) yield three different results (Fig.~\ref{fig:trilateration}).
The intersection of accounts by 
${\cal A}$
and ${\cal B}$ 
(``Pair 1") 
would locate $L$ very inaccurately around the airport, west of $E$, and is incompatible with the observation that the missile was heading from South to North. 
The intersection of accounts by 
${\cal C}$
and 
${\cal D}$
(``Pair 2")
would locate $L$ 
south-west of $E$. 
The intersection of accounts by 
${\cal E}$ 
and ${\cal F}$
(``Pair 3")
would locate $L$ 
south-east of $E$. 
We retain that the former position (``Pair 1") is much less likely than both latter ones (``Pair 2", ``Pair 3").

\begin{figure}[t]
\centering
\showfigures{\includegraphics[width=1\textwidth]{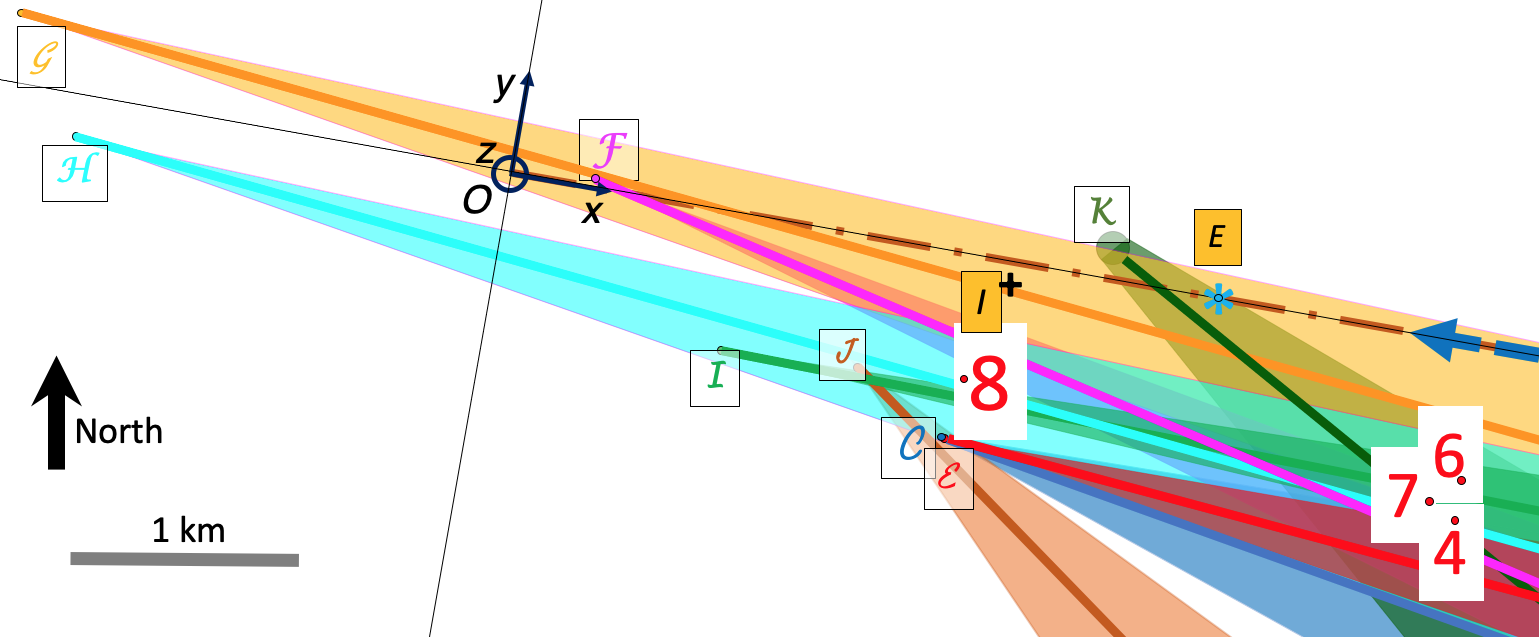}}\\
\caption{{\bf Triangulation for the missile launch site $L$ based on accounts of direction.}
Same caption as Fig.~\ref{fig:synthesis}.
The directions reported by the witnesses for $L$ are indicated by straight lines and sectors, reflecting the uncertainties.
The confidence interval for  ${\cal K}$'s position is indicated by the green circle. 
The points labeled with numbers are the possible launch site location as determined from our minimization procedure (see text).
}
\label{fig:triangulation}
\end{figure}

\begin{figure}[t]
\centering
\showfigures{\includegraphics[width=0.6\textwidth]{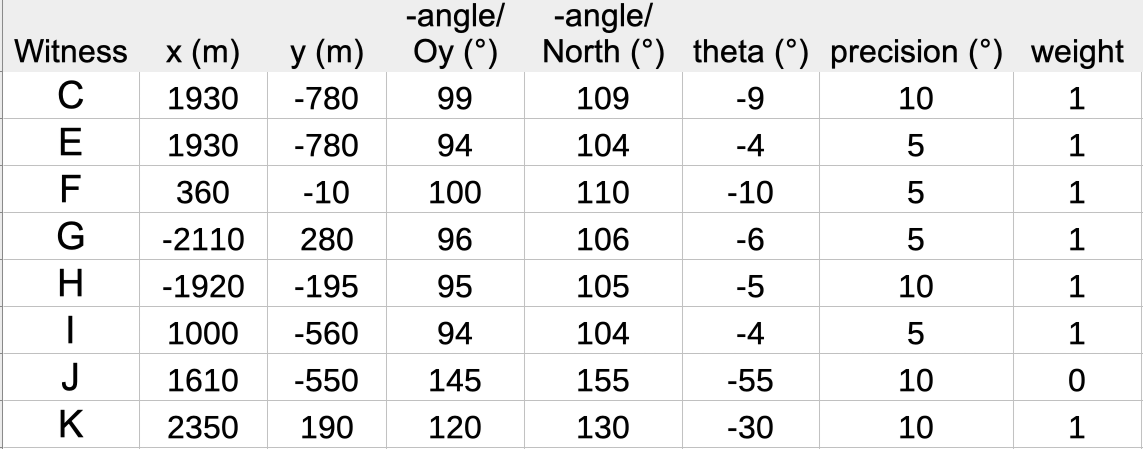}}\\
\showfigures{\includegraphics[width=0.8\textwidth]{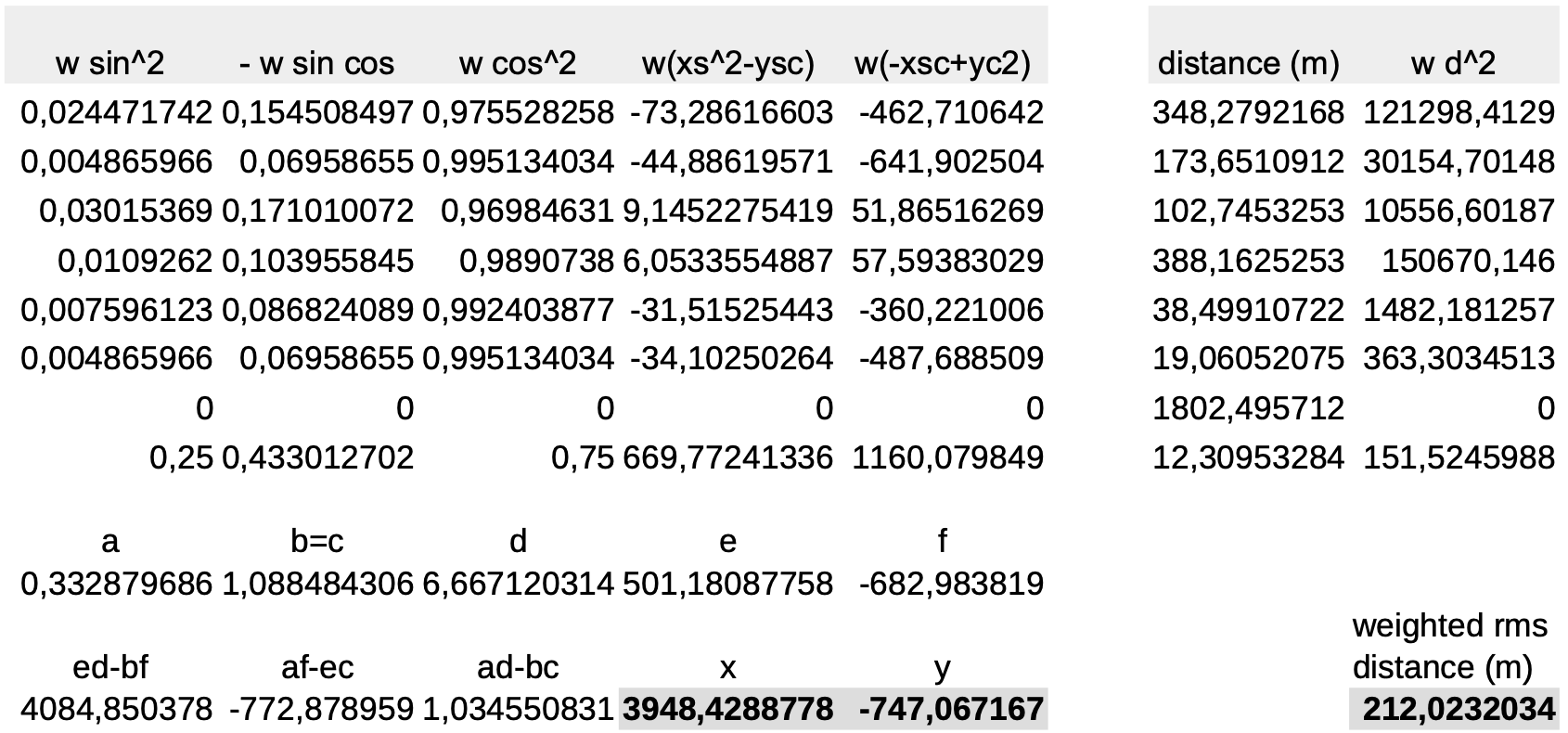}}\\
\caption{{\bf Example of triangulation data.}
Top: data from 8 witness accounts. Bottom: step by step decomposition of the minimisation calculation, without rounding. In this example, all weights are equal ($w=1$), except for one witness which is removed ($w=0$). Abbreviations: ``s" = sin, ``c" = cos. }
\label{data_triangulation}
\end{figure}

\section{Triangulation, using directions}
\label{sec:supp_triangulation_directions}

The position of the missile launch site can also be triangulated using witness accounts of the direction from which the missile was observed to originate.
 Some of them extrapolate the (straight or curved) part of the trajectory that they saw.
We have identified 8 accounts which can be exploited
(Fig.~\ref{fig:triangulation}).
Based on the witness statements (size and distance of the landmarks they indicate, quality of their drawings) we estimate that four of them 
(${\cal E}$,
${\cal F}$,
${\cal G}$,
${\cal I}$)
give a direction that has an uncertainty of about  $\pm 5^\circ$
and  four others 
(${\cal C}$,
${\cal H}$,
${\cal J}$,
${\cal K}$)
 of about  $\pm 10^\circ$ (Fig.~\ref{data_triangulation}, top).
Consider a witness  $k$ ($k=1$ to 8) located in the $Oxy$ plane at $W_k (x_k,y_k)$ and who reported the launch site $L$ to be in a direction $\vec{u}_k = (\cos \theta_k, \sin \theta_k)$. 
The eight rays defined by  each $W_k$ and $\vec{u}_k$ do not  intersect at a single point; finding $L$ by triangulation is thus  an overdetermined problem (Fig.~\ref{fig:triangulation}). 
We suggest to minimize the square distances  $d_k$ of a point $(x,y)$ to each of the rays: 
\begin{eqnarray}
d_k^2 &=& [ - (x-x_k) \sin \theta_k + (y-y_k) \cos \theta_k ]^2 \nonumber \\
&=& (x-x_k)^2 \sin^2 \theta_k - 2 (x-x_k) (y-y_k) \sin \theta_k \cos \theta_k + (y-y_k)^2 \cos^2 \theta_k
\end{eqnarray}
We look for the minimum of the root mean square (r.m.s.) distance:
\begin{equation}
\bar{d^2} = \frac{\sum_{k=1}^8 w_k d_k^2} {\sum_{k=1}^8 w_k}
\label{eq:def_d2}
\end{equation}
Here we introduce the possibility of using weights $w_k$. For instance,  the weighting method could enforce that
 an observer who is near the determined launch site should have a smaller acceptable range of deviation $d$ compared to an observer with the same angular uncertainty who is farther from the determined launch site. Alternatively, 
we assign weights proportional to the inverse of the confidence interval size so that
the four most accurate witnesses are given weights twice as large as those of the four less accurate ones.

Minimizing $\bar{d^2}$ (Eq.~(\ref{eq:def_d2})) amounts to solving the two equations $ \partial_x \bar{d^2}  = \partial_y \bar{d^2} = 0$, i.e.:
\begin{eqnarray}
\sum_{k=1}^8 w_k \left[ 2 (x-x_k) \sin^2 \theta_k - 2 (y-y_k) \sin \theta_k \cos \theta_k\right] &=& 0 
\label{eq:linear_optimization_a} \\
\sum_{k=1}^8 w_k \left[  - 2 (x-x_k) \sin \theta_k \cos \theta_k + 2 (y-y_k) \cos^2 \theta_k\right] &=& 0,
\label{eq:linear_optimization_b}
\end{eqnarray}
which yields
only one solution.

Weights of 2 for the four precise witnesses and of 1 for the four  others
yield $x  = + 2092$~m, $y = -480$~m, with a  r.m.s. distance $\sqrt{\bar{d^2}} = 328$~m. 
This  point (marked by a number ``8" in Fig.~\ref{fig:triangulation}), is compatible with Pair 2 of distance accounts (Section \ref{sec:triangulation_sound_speed}).
Since it is incompatible with five out of eight direction accounts
(${\cal C}$,
${\cal E}$,
${\cal I}$,
${\cal J}$,
${\cal K}$),
it is interesting to test how robust this result is with respect to the  selection of accounts.

Choosing the method to remove outliers has an impact on the result. 
One possibility is to minimize a norm other than the r.m.s. distance  $\bar{d^2}$ (Eq.~(\ref{eq:def_d2})); for instance the average distance $\bar{d}$.
For simplicity, we use here the Least Trimmed Squares method~\cite{Rousseeuw}, which consists in removing one by one the data points in order of importance. 
To facilitate the intuitive understanding of the method and its effect, we adapt it manually as follows.

If we consider
${\cal J}$
 as the most distant outlier and exclude it,  the resulting position and r.m.s. distance are more consistent and robust to changes in the weights.
For instance, with all weights equal,  $x = 3948$~m, $y = -747$~m, while with $w=2$ for the most precise ones we find $x = 3967$~m, $y = - 722$~m. 
The r.m.s. distance is around 212~m in both cases (Fig.~\ref{data_triangulation}, bottom).
This  point is compatible with the observations of all 7 out of 7  accounts (marked ``7" in Fig.~\ref{fig:triangulation}).
Removing one by one  accounts which are redundant and/or barely discriminating
 does not significantly affect the result, which is dominated by the crossing point between 
 ${\cal E}$ 
and  ${\cal K}$
 (Fig.~\ref{fig:triangulation}).
More generally, by testing both weighted and unweighted combinations, we aggregate the results for the launch position $L$ 
under the values $x_L = + 4080 \pm 120$~m, $y_L = -700 \pm 100$~m, with a r.m.s. residual of $\sqrt{\bar{d^2}} \approx 200$~m (Fig.~\ref{fig:synthesis}). 
This is
compatible with  Pair 3
of distance accounts (Section \ref{sec:triangulation_sound_speed}).

In summary, this Section shows the inference method, based on aggregating diverse information; 
as opposed to the deductive method, starting with a closed list of candidate points and examining their compatibility with constraints.
It also evidences how outlier trimming can strongly affect the results.

\section{Missile launchers}
\label{sec:supp_missile_launchers}

A missile launcher is a tube that contains the missile before and during the launch. 
Several witnesses mentioned that a pair of missile launchers has been found.
Some other witnesses diverged on the existence of these launchers, their number,
or their
finding date  (Ref.~\cite{mutsinzi} p. 153-155).

We use a  detailed map at scale 1/2500 established by French experts (Ref.~\cite{oosterlinck} p.~60)
and exploited two accounts which were  explicit enough.
One was a map 
with the place where launchers were supposed to have been found
(Ref.~\cite{bruguiere}, 8124);
from this map we infer 
$x = +3800 \pm 200$~m, $y = -1200 \pm 100$~m.
The other originates from Rwandan investigators (Ref.~\cite{mutsinzi} p. 154, 156, 157) who,
while doubting the finding was real, mentioned it would have been located  in a marsh, and provided distances to two landmarks.
From this information,
 we infer  two positions
 only the west one being near a marsh: it is located at  $x = +4100 \pm 100$~m, $y = - 1120 \pm 40$~m. 
Together, these informations would suggest that the missile launchers have been found around $x = +4000 \pm 100$~m, $y = - 1130 \pm 30$~m (green cross on Fig.~\ref{fig:synthesis}). 
This is on dry land, near a marsh but not in it, and corresponds to $z=1334\pm 2$~m.
\label{sec:launch_pos_value}

\section{Two-phase missile trajectory duration}
\label{sec:supp_miss_duration}

 We compare the durations of the missile trajectory  and of its visibility.  For simplicity, we approximate the missile trajectory  with two phases. First, a propulsion phase with constant acceleration $a_p$, during which the engine flames make the trajectory visible, increases the missile speed from 0 to $v_p$. 
 Its duration  $t_p$  and length  $\ell_p$ are related by the four equivalent equations:
 \begin{equation}
v_p = a_p t_p  \; \; \; (a)
; \; \; \; \; \; \;
2 \ell_p = a_p t_p^2  \; \; \; (b)
; \; \; \; \; \; \;
2 a_p \ell_p =  v_p^2  \; \; \; (c)
; \; \; \; \; \; \;
2 \ell_p = v_p t_p  \; \; \; (d).
\label{eq:accelerated_phase}
\end{equation}
Then a  non propulsion phase at constant speed $v_p$ during which the trajectory is less visible. Its duration $t_{np}$ and length $\ell_{np}$ obey:
\begin{equation}
\ell_{np} = v_p t_{np}.
\label{eq:non_accelerated_phase}
\end{equation}
The total missile trajectory has  duration $t_m$ and length $\ell_m$: 
\begin{equation}
\ell_m = \ell_p + \ell_{np} \; \; \; (a)
\label{eq:total_two_phases}
; \; \; \; \; \; \;
t_m =  t_p + t_{np} \; \; \; (b).
\end{equation}
The various parameters for the three different missile types are given in Table~\ref{table:missile_trajectory_duration}: the publicly available information includes either the duration $t_p$ of the propulsion phase, its acceleration $a_p$ or its final speed $v_p$. We use this information to deduce the values of other parameters.

\begin{table}[t]
\centering
\caption{ 
{\bf Missile trajectory phases.}
Quantitative description of the propulsion and non propulsion phases, for three missile types. 
The non propulsion phase corresponds  to a scenario with a missile launched from South-East of the interception location.
The  confidence interval of the total missile trajectory length 
$\ell_m = \modifxE{1559} \pm \modifxE{362}$~m
propagates to $\ell_{np}$, $t_{np}$ and $t_m$; in the case of the Stinger, the non-propulsion phase can be non-existent, which results in an asymmetric confidence interval represented here as (minimum; mean; maximum).
}
\begin{ruledtabular}
\begin{tabular}{rcccc}
Quantity & SAM & Stinger &  Mistral \\
\hline%
\hline
Propulsion duration  $t_p$ 
& 2 s &  3.49 s & 2.6 s \\
    & Ref.~\cite{cranfield} p.~54 &  Eq.~(\ref{eq:accelerated_phase}a) & Ref.~\cite{mindef_mistral} \\
\hline
Propulsion length  $\ell_p$  
& 686 m &  1316 m & 1114 m  \\
  & Eq.~(\ref{eq:accelerated_phase}d) &  Eq.~(\ref{eq:accelerated_phase}c) & Eq.~(\ref{eq:accelerated_phase}d) \\
\hline
Speed at end of propulsion  $v_p$   
& 686 m/s &  754 m/s & 857 m/s  \\
    & Ref.~\cite{cranfield} p.~54 &  Ref.~\cite{oosterlinck} p. 162-167 & Ref.~\cite{mindef_mistral} \\
\hline
Propulsion acceleration   $a_p$  
& 343  m/s$^2$  &  216  m/s$^2$  & 330  m/s$^2$  \\ 
   & Eq.~(\ref{eq:accelerated_phase}a) &  Ref.~\cite{fas_man} p. 3 & Eq.~(\ref{eq:accelerated_phase}a) \\
\hline
Non-propulsion length  $\ell_{np}$  
&   $\modifxE{873} \pm \modifxE{362}$ m &  (0 m; \modifxE{243} m; \modifxE{605} m) & $\modifxE{445} \pm \modifxE{362}$ m  \\
     & Eq.~(\ref{eq:total_two_phases}a) &  Eq.~(\ref{eq:total_two_phases}a) & Eq.~(\ref{eq:total_two_phases}a) \\
\hline
Non-propulsion duration  $t_{np}$  
& $\modifxE{1.27} \pm \modifxE{0.53}$ s &  (0 s; \modifxE{0.32} s; \modifxE{0.8} s)  &  $\modifxE{0.52} \pm \modifxE{0.42}$ s   \\
   & Eq.~(\ref{eq:non_accelerated_phase}) &  Eq.~(\ref{eq:non_accelerated_phase}) & Eq.~(\ref{eq:non_accelerated_phase}) \\
\hline
Total duration  $t_m$ 
& $\modifxE{3.27} \pm \modifxE{0.53}$ s &  (\modifxE{3.32} s; \modifxE{3.81}; \modifxE{4.29} s) & $\modifxE{3.2} \pm \modifxE{0.42}$ s \\
   & Eq.~(\ref{eq:total_two_phases}b) &  Eqs.~(\ref{eq:accelerated_phase}b,\ref{eq:total_two_phases}b) & Eq.~(\ref{eq:total_two_phases}b) \\
\end{tabular}
\end{ruledtabular}
\label{table:missile_trajectory_duration}
\end{table}

\clearpage

\begin{table}[t]
\centering
\caption{\bf Results, notations, uncertainties and sources.} 
We present here excess digits; for legibility, asymmetric confidence intervals (e.g. for ${\dot{z}}_I$, Section~\ref{sec:impact_angle_orientation}) have been symmetrized.
 To test the sensitivity to rounding (Section~\ref{sec:sensitivity}), values calculated with $x_E = 3000 \pm 300$~m are displayed as tiny figures in italics within brackets.
The {\it Missile before interception} part corresponds  to a scenario with a missile launched from South-East of the interception location.
\begin{ruledtabular}
\begin{tabular}{| r l r l l  p{5cm} |}
Name  & {Symbol}
&   Value  & uncertainty  & Section & Source  \\
\hline%
\hline
{\it Aircraft before interception} &  &  & &  &  \\
\hline
Time at  {crew} announcement &  $t_A$ & 20:21:27  &  $\pm$ 1 s &  \ref{sec:interception_description} & Ref.~\cite{bande} p.~11  \\
 Position at  {crew} announcement & $x_A$  & 37 0{40} m  &  $\pm$ 1 852 m &   \ref{sec:interception_description} & Ref.~\cite{bande} p.~11    \\
\phantom{Posi}"\phantom{ion a}"\phantom{ annou}"\phantom{cement} & $z_A$  & 3 657 m  &  $\pm$ 300 m &   \ref{sec:interception_description} & Eq.~(\ref{eq:glide_slope_b})   \\
 {Velocity}  at  {crew} announcement   & ${\dot{x}}_A$
  & $-181.5$ m/s  &  $\pm$ 10 m/s & \ref{sec:speed_at_impact} & Eqs~(\ref{eq:speed_position_vs_time_a}-\ref{eq:speed_profile})   \\
Average deceleration & $a$  & 0.397~m/s$^2$  &  $\pm$ 0.012~m/s$^2$ &  \ref{sec:speed_at_impact}  & Eqs~(\ref{eq:speed_position_vs_time_a}-\ref{eq:speed_profile})   \\
Glide slope angle {in vertical plane} &  $\gamma$  &  3$^\circ$ &  $\pm$ 0.01$^\circ$  &  \ref{sec:airport}  &  Ref.~\cite{cranfield} p.~106 \\
Position at glide interception & $x_G$  &  14 4{46} m &  $\pm$ 185~m  &  \ref{sec:airport}  &  Ref.~\cite{cranfield} p.~106  \\
\phantom{Posi}"\phantom{ion a}"\phantom{ gli}"\phantom{de inter}"\phantom{eption}
 & $y_G$  &  0 m &  $\pm$ 125~m  &  \ref{sec:fall}  &  Eq.~(\ref{eq:glide_slope_a})   \\
\phantom{Posi}"\phantom{ion a}"\phantom{ gli}"\phantom{de inter}"\phantom{eption}
& $z_G$  &  750 m &  $\pm$ 125~m  &  \ref{sec:fall}  &  Eq.~(\ref{eq:glide_slope_b})   \\
Time at glide interception & $t_G$  &  20:23:56  &  $\pm$ 10 s  &  \ref{sec:speed_at_impact}  &  by-product of Eqs~(\ref{eq:speed_position_vs_time_a}-\ref{eq:speed_profile}) \\
 {Velocity}  at glide interception & ${\dot{x}}_G$
 &  $-122.5$ m/s &  $\pm$ 10 m/s &  \ref{sec:speed_at_impact}  &   by-product of Eqs~(\ref{eq:speed_position_vs_time_a}-\ref{eq:speed_profile}) \\
\hline
\hline
{\it Missile before interception} &  &  & &  &  \\
\hline
Missile number & $$ & 2  &  $\pm$ 1 &   \ref{sec:missile_number} &  witnesses \\
Launch position
& $x_L$ & 4 000 m  &  $\pm$ 100 m &   \ref{sec:launch_pos_value} & witnesses with triangulation  \\
 \phantom{Laun}" \phantom{h pos}" \phantom{tion} 
& $y_L$ & $-1$ 130 m  &  $\pm$ 30 m &   \ref{sec:launch_pos_value} &  witnesses with triangulation  \\
 \phantom{Laun}" \phantom{h pos}" \phantom{tion} 
& $z_L$ &  1 334 m &  $\pm$ 2 m &   \ref{sec:launch_pos_value} &  Ref.~\cite{oosterlinck} p.~60 \\
 {Launch angle in horizontal plane} & $$ & \modifxE{132}$^\circ$  {\tiny \it (130)}  &  $\pm$  \modifxE{11}$^\circ$ {\tiny \it (10)} 
 &   \ref{sec:trajectory_orientation} &  geometry    \\
  {Launch angle in vertical plane} & $\varphi$ & \modifxE{36.3}$^\circ$  {\tiny \it (36)}  &  $\pm$ \modifxE{12.3}$^\circ$ {\tiny \it (11)} 
 &   \ref{sec:trajectory_orientation} & Eq.~(\ref{eq:phi_alpha_psi})   \\
 {Average angle  in vertical plane} & $$ &    \modifxE{12}$^\circ$ {\tiny \it (12)}
 &   $\pm$ \modifxE{4}$^\circ$ {\tiny \it (3)}
 &   \ref{sec:trajectory_orientation} &  geometry  \\
Trajectory length & ${\ell_m}$ &   \modifxE{1 559} m {\tiny \it (1 540)}  &  $\pm$  \modifxE{362} m {\tiny \it (316)} 
&   \ref{sec:trajectory_orientation} &  geometry   \\
Trajectory duration: 
 if SAM   & ${t_m}$ & \modifxE{3.27} s {\tiny \it (3.25)} &  $\pm$ \modifxE{0.53} s  {\tiny \it (0.44)}  
 & \ref{sec:missile_type}    & 
   {Table~\ref{table:missile_trajectory_duration}} \\
\phantom{Traj}"\phantom{ctory dur}"\phantom{tion} 
 if Stinger & {"} &  \modifxE{3.8} s {\tiny \it (3.79)} &  $\pm$ \modifxE{0.49} s {\tiny \it (0.4)}  
 & \ref{sec:missile_type}    &    {Table~\ref{table:missile_trajectory_duration}} \\
\phantom{Traj}"\phantom{ctory dur}"\phantom{tion} 
 if Mistral & {"}  &  \modifxE{3.2} s {\tiny \it (3.1)} &  $\pm$ \modifxE{0.42} s {\tiny \it (0.35)} 
 & \ref{sec:missile_type}    &    {Table~\ref{table:missile_trajectory_duration}} \\
\hline
\hline
{\it Aircraft-missile interception} &  &  & &  &  \\
\hline
Position  {of} interception & $x_E$ 
&  \modifxE{2 970} m {\tiny \it (3 000)}&  $\pm$ \modifxE{345}  m {\tiny \it (300)} 
&   \ref{sec:pos_and_speed_interception}  & geometry and witnesses  \\
\phantom{Posi}"\phantom{ion a}"\phantom{ enco}"\phantom{nter} 
& $y_E$ & 0 m  &  $\pm$ 20 m &  \ref{sec:pos_and_speed_interception}   & Eq.~(\ref{eq:glide_slope_a})   \\
\phantom{Posi}"\phantom{ion a}"\phantom{ enco}"\phantom{nter} 
& $z_E$ & \modifxE{1 639} m {\tiny \it (1 641)}
&  $\pm$ \modifxE{31} m {\tiny \it (30)}
&  \ref{sec:pos_and_speed_interception}   & Eq.~(\ref{eq:glide_slope_b})   \\
 Aircraft's velocity at interception & ${\dot{x}}_E$  &  $-77.1$ m/s &  $\pm$  10 m/s &  \ref{sec:pos_and_speed_interception}  & Eq.~(\ref{eq:speed_profile})  \\
\phantom{Airc}"\phantom{aft velo}"\phantom{ity a}"\phantom{ enco}"\phantom{nter} 
& ${\dot{y}}_E$ & 0 m/s  &  $\pm$ 0.5 m/s & \ref{sec:pos_and_speed_interception}   & Eq.~(\ref{eq:glide_slope_a})   \\
\phantom{Airc}"\phantom{aft velo}"\phantom{ity a}"\phantom{ enco}"\phantom{nter} 
& ${\dot{z}}_E$ & $-4$ m/s  &  $\pm$ 0.5 m/s & \ref{sec:pos_and_speed_interception}   & Eq.~(\ref{eq:glide_slope_b})   \\
\hline
\hline
{\it Aircraft  {impact on ground}} &  &  & &  &  \\
\hline
Time of interception and impact & $t_E, t_I$ & 20:26:01  &  $\pm$ 11 s &  \ref{sec:interception_description}   &  Ref.~\cite{bande} p.~12, 13  \\
Fall duration {$t_I-t_E$} &  $D$
& 11 s  &  $\pm$ 2.5 s & \ref{sec:pos_and_speed_interception}  & geometry and witness  \\
Fall vertical acceleration & ${\ddot{z}}$ & $-3.08$~m/s$^2$  &  $\pm$ 1 ~m/s$^2$ & \ref{sec:impact_angle_orientation}  &  geometry and velocity \\
Position at impact
& $x_I$ & 2 160 m  &  $\pm$ 20 m & \ref{sec:interception_description}  & Ref.~\cite{oosterlinck} p.~188, 192  \\
\phantom{Posi}"\phantom{ion a}"\phantom{ imp}"\phantom{ct} 
& $y_I$ & $-100$ m  &  $\pm$ 20 m &  \ref{sec:interception_description} &  Ref.~\cite{oosterlinck} p.~188, 192  \\
\phantom{Posi}"\phantom{ion a}"\phantom{ imp}"\phantom{ct} 
& $z_I$ &  1 410 m &  $\pm$ 2 m & \ref{sec:interception_description}  & Ref.~\cite{oosterlinck} p.~188, 192  \\
 {Velocity} at impact & ${\dot{x}}_I$ &  $-72.7$ m/s &  $\pm$ 10 m/s & \ref{sec:speed_at_impact}  &  Eq.~(\ref{eq:speed_profile}) \\
\phantom{Velo}"\phantom{ity a}"\phantom{ imp}"\phantom{ct} 
& ${\dot{y}}_I$ &  -18.3 m/s &  $\pm$ 2.3 m/s &   \ref{sec:impact_angle_orientation}  &  geometry and velocity   \\
\phantom{Velo}"\phantom{ity a}"\phantom{ imp}"\phantom{ct} 
& ${\dot{z}}_I$ &  \modifxE{-37.7} m/s {\tiny \it (38)} &  $\pm$ \modifxE{7.2} m/s {\tiny \it (7)} 
& \ref{sec:impact_angle_orientation}  &  geometry and velocity   \\
Angle  at impact 
 {in vertical plane}   & $$ & \modifxE{26.1}$^\circ$ {\tiny \it (26.2)}  &  $\pm$ \modifxE{4.8}$^\circ$ {\tiny \it (5)}
 &    \ref{sec:impact_angle_orientation}  &  geometry  \\
\phantom{Angle}"\phantom{t i}"\phantom{pa}"\phantom{t} 
 {in horizontal plane}  & $ {\beta}$ & {13.7$^\circ$}  
 &  $\pm$ $4^\circ$ & \ref{sec:lateral_fall}  & Ref.~\cite{oosterlinck} p.~189  \\
\end{tabular}
\end{ruledtabular}
\label{table:results}
\end{table}


\begin{thebibliography}{0}%
\makeatletter
\providecommand \@ifxundefined [1]{%
 \@ifx{#1\undefined}
}%
\providecommand \@ifnum [1]{%
 \ifnum #1\expandafter \@firstoftwo
 \else \expandafter \@secondoftwo
 \fi
}%
\providecommand \@ifx [1]{%
 \ifx #1\expandafter \@firstoftwo
 \else \expandafter \@secondoftwo
 \fi
}%
\providecommand \natexlab [1]{#1}%
\providecommand \enquote  [1]{``#1''}%
\providecommand \bibnamefont  [1]{#1}%
\providecommand \bibfnamefont [1]{#1}%
\providecommand \citenamefont [1]{#1}%
\providecommand \href@noop [0]{\@secondoftwo}%
\providecommand \href [0]{\begingroup \@sanitize@url \@href}%
\providecommand \@href[1]{\@@startlink{#1}\@@href}%
\providecommand \@@href[1]{\endgroup#1\@@endlink}%
\providecommand \@sanitize@url [0]{\catcode `\\12\catcode `\$12\catcode
  `\&12\catcode `\#12\catcode `\^12\catcode `\_12\catcode `\%12\relax}%
\providecommand \@@startlink[1]{}%
\providecommand \@@endlink[0]{}%
\providecommand \url  [0]{\begingroup\@sanitize@url \@url }%
\providecommand \@url [1]{\endgroup\@href {#1}{\urlprefix }}%
\providecommand \urlprefix  [0]{URL }%
\providecommand \Eprint [0]{\href }%
\providecommand \doibase [0]{http://dx.doi.org/}%
\providecommand \selectlanguage [0]{\@gobble}%
\providecommand \bibinfo  [0]{\@secondoftwo}%
\providecommand \bibfield  [0]{\@secondoftwo}%
\providecommand \translation [1]{[#1]}%
\providecommand \BibitemOpen [0]{}%
\providecommand \bibitemStop [0]{}%
\providecommand \bibitemNoStop [0]{.\EOS\space}%
\providecommand \EOS [0]{\spacefactor3000\relax}%
\providecommand \BibitemShut  [1]{\csname bibitem#1\endcsname}%
\let\auto@bib@innerbib\@empty
\end{thebibliography}%


\clearpage





\begin{thebibliography}{99}






\bibitem{bruguiere}
{\it Instruction concernant l'attentat du 6 avril 1994 \`a Kigali}, Tribunal de Grande Instance de Paris, 15 February 2022 (in French), \\
\url{<https://francegenocidetutsi.fr/fgtarchives.php>}

\bibitem{belgian_investigators}
Smeets, P. and Paque, J. 
{\it Rapport d'enqu\^ete, Sinistre a\'erien du 06 avr 94 \`a Kigali - Falcon 50}, Note Auditorat Militaire N$^\circ$ 02.02545W94/Cab 8,
1st August 1994 (partly in French and partly in Dutch),\\
 \url{<https://francegenocidetutsi.fr/fgtarchives.php>}, 7154.

\bibitem{bande}
Plantin de Hugues, P. 
{\it Rapport d'expertise}, Tribunal de Grande Instance de Paris,
10 April 2002 (in French),\\
 \url{<https://francegenocidetutsi.fr/fgtarchives.php>}, 6036

\bibitem{cranfield}
Warden, M.C. and McClue, W.A.
{\it Investigation into the crash of Dassault Falcon 50 registration number 9XR-NN on 6 April 1994 carrying former President Juvenal Habyarimana},
 Defence Academy of the United Kingdom and Cranfield University, 

 27 February 2009
    (exists in english and in French; page numbers here refer to the English version), \\
 \url{<https://francegenocidetutsi.fr/documents/cranfield-en.pdf>}
 
\bibitem{mutsinzi}
   Mutsinzi, J.  {\it et al.}, {\it Report of the Investigation into the Causes and Circumstances of and Responsibility for the Attack of 06/04/1994 against the Falcon 50 Rwandan Presidential Aeroplane, Registration number 9XR-NN}, Republic of Rwanda, 20 April 2009  (exists in three languages; page numbers here refer to the english version),\\
  \url{<http://mutsinzireport.com>}
 

\bibitem{mutsinzi_video}
  Mutsinzi, J.  {\it et al.}, 
 
{\it Video Analysis of Habyarimana Plane Crash}, 11 January 2010,\\
\url{<https://www.youtube.com/watch?v=0bRJbPL1d3Y>}
  
\bibitem{oosterlinck}
  Oosterlinck, C.
   {\it et al.},
  
   {\it Rapport d'expertise. Destruction en vol du Falcon 50 Kigali (Rwanda)}, Tribunal de Grande Instance de Paris, 5 January 2012  (in French),\\
   \url{<http://francegenocidetutsi.org/rapport-balstique-attentat-contre-habyarimana-6-4-1994.pdf>}

\bibitem{acoustic}
Serre, J.-P. {\it Rapport compl\'ementaire en acoustique},
Tribunal de Grande Instance de Paris, 
 3 January 2012 (in French),\\
 \url{<https://francegenocidetutsi.fr/documents/RapportExpertAcoustiqueJanvier2012.pdf>}\\
Annexes A to E,  \\
\url{<https://francegenocidetutsi.fr/documents/AnnexesRapportAcoustique05012012.pdf>}



\bibitem{FSF}
Flight Safety Foundation {\it Approach-and-landing Accident Reduction Tool Kit - Briefing Note 7.1 - Stabilized Approach}, p. 134, table 1, point 8,\\
\url{<https://flightsafety.org/wp-content/uploads/2016/09/alar_bn7-1stablizedappr.pdf>}

\bibitem{CANSO}
Civil Air Navigation Services Organisation (CANSO), {\it Unstable Approaches ATC Considerations}, Appendix A, p. 15,\\
\url{<https://www.icao.int/safety/RunwaySafety/Documents/Unstable%20Approaches-ATC%20Considerations.pdf>}



\bibitem{eurocontrol}
Eurocontrol Aviation Learning Centre, {\it Aircraft Performance Database (only indicative) - FA50 Dassault-Breguet T16},\\
 \url{<https://contentzone.eurocontrol.int/aircraftperformance/details.aspx?ICAO=FA50&ICAOFilter=FA50>}

\bibitem{wp_falcon} 
Dassault Falcon 50, wikipedia  (in French), \\
\url{<https://fr.wikipedia.org/wiki/Dassault_Falcon_50>}

\bibitem{mindef_mistral}
SIRPATerre, {\it Poste de tir Mistral - caract\'eristiques techniques},  Arm\'ee de terre, 2012  (in French),\\
 \url{<www.defense.gouv.fr/terre/.pdf>} 

\bibitem{wp_RDX} 
RDX, wikipedia,\\
\url{<https://en.wikipedia.org/wiki/RDX>}


\bibitem{RapportEvitement}
Oosterlinck, C.
   {\it et al.},
  
{\it Rapport d'expertise. Destruction en vol du Falcon 50 Kigali (Rwanda)}, Tribunal de Grande Instance de Paris, Compl\'ement de mission (Manoeuvre d'\'evitement). 10 May 2013  (in French), with an Annex,\\
   \url{<http://francegenocidetutsi.org/RapportExpertEvitement10mai2013.pdf>}\\
  
   \url{<http://francegenocidetutsi.org/AnnexesRapportEvitement10mai2013.pdf>}


\bibitem{Hick}
Hick, W.E., ``On the rate of gain of information", 
{\it Quart. J. Exp. Psychol.} 
4, 11-26 (1952).\\
\url{<http://www2.psychology.uiowa.edu/faculty/mordkoff/InfoProc/pdfs/Hick}~\url{1952.pdf>}

\bibitem{Akdag_master}
Akdag, R., ``Evaluation of fighter evasive maneuvers against proportional navigation missiles", Master thesis, Turkish naval academy, Istanbul (2005)\\
\url{<https://web.itu.edu.tr/altilar/thesis/msc/AKDAG05.pdf>}


\bibitem{simbad}
Matra D\'efense, {\it Simbad: Syst\`eme Int\'egr\'e Mistral Bi-munitions pour l'Autod\'efense} (in French),\\
\url{<http://www.netmarine.net>} (consulted August 17th, 2014; no longer online). 


\bibitem{ancel_nepassubir}
Ancel, G., {\it Rapport d'expertise sur l'assassinat du pr\'esident Habyarimana, le 6 avril 1994}, blog, 28 January 2018  (in French),\\
 \url{<https://nepassubir.fr/2018/01/28/>}  





\bibitem{rcaa}
Rwanda Civil Aviation Authority,
{\it HRYR Aerodrom chart ICAO - Kigali, Rwanda}, 25 January 2024,\\
\url{<https://rac.co.rw/eAIP%20Rwanda/1ST%20EDITION_2024_01_25>}

\bibitem{manuel}
Dassault Aviation, {\it Falcon 50 - Airplane Flight Manual},\\
\url{<https://www.avialogs.com/aircraft-d/dassault/item/2743-falcon-50-airplane-flight-manual>}


\bibitem{Rousseeuw}
{Rousseeuw, P. J. and Leroy, A. M., {\it Robust Regression and Outlier Detection}, Wiley (2005),\\
\url{<https://doi.org/10.1002/0471725382>}
}


\bibitem{fas_man}
Federation of American Scientists - Military Analysis Network,
{\it MCWP 3-25.10 - Chapter 2 - Stinger Weapon System},\\ 
 \url{<https://man.fas.org/dod-101/sys/land/docs/ch2.pdf>}
















 


\end{thebibliography}
\end{document}